\documentclass[12pt]{article}

\usepackage[latin1]{inputenc}   
\usepackage[francais] {babel}    
\usepackage{graphicx}       
\usepackage{amsmath}        

\newcommand{\be}{\begin{equation}}
\newcommand{\ee}{\end{equation}}
\newcommand{\bc}{\begin{center}}
\newcommand{\ec}{\end{center}}
\newcommand{\glv}{\gamma_{LV}}
\newcommand{\gsl}{\gamma_{SL}}
\newcommand{\gsv}{\gamma_{SV}}

\begin{document}

\bc
 \textbf {Intrusion and extrusion of water in hydrophobic mesopores}\\
 \vspace{1cm}
        B. LEFEVRE$^{a}$, A. SAUGEY$^{b,c}$, J.L. BARRAT$^{d}$,
        L. BOCQUET$^{d}$, E. CHARLAIX$^{d}$, P.F. GOBIN$^{b}$ and G. VIGIER$^{b}$
        .\\

 \vspace{1cm}

 $^{a}$    \textit{Laboratoire de Mat\'eriaux Catalytiques et Catalyse en Chimie Organique,
            8, rue de l'Ecole Normale, 34296 Montpellier Cedex 05, France}\\
 $^{b}$     \textit{Groupe d'Etudes de M\'etallurgie Physique et de
            Physique des Mat\'eriaux, 20, Avenue Albert
            Einstein, 69621 Villeurbanne Cedex, France}\\
 $^{c}$     \textit{Laboratoire de Tribologie et Dynamique des Syst\`emes, Ecole Centrale de Lyon,
            36 Avenue Guy de Collongues, BP163, 69134 Ecully Cedex, France}\\
 $^{d}$     \textit{Laboratoire de Physique de la Mati\`ere Condens\'ee et Nanostructures,
            Universit\'e Claude Bernard, 6 rue Amp\`ere, 69622 Villeurbanne Cedex, France}\\
 $^{*}$     \textit{\\}

\centerline{\bf ABSTRACT}

\ec

 {\bf We present experimental and theoretical results on
intrusion-extrusion cycles of water in hydrophobic mesoporous
materials, characterized by independent cylindrical pores. The
intrusion, which takes place above the bulk saturation pressure,
can be well described using a macroscopic capillary model. Once
the material is saturated with water, extrusion takes place upon
reduction of the externally applied pressure; Our results for the
extrusion pressure can only be understood by assuming that the
limiting  extrusion mechanism is the nucleation of a vapour bubble
inside the pores. A comparison of calculated and experimental
nucleation pressures shows that a proper inclusion of line tension
effects is necessary to account for the observed values of
nucleation barriers. Negative line tensions of order $10^{-11}
\mathrm{J.m}^{-1}$ are found for our system, in reasonable
agreement with other experimental estimates of this quantity.}

\section{Introduction}

Porous materials are involved in many industrial processes such as
catalysis, filtration, chromatography, etc...  In order to
understand such  processing technologies, accurate information on
the porous texture (surface area, pore size distribution and pore
shape) is needed.  The most widely used characterization methods
are based on adsorption isotherms and capillary condensation,
usually described as a gas-liquid phase transition shifted by
confinement \cite{Israelachvili}. Other methods, based on
capillary evaporation (mercury or water porosimetry on hydrophobic
porous materials) often provide a useful alternative. Experimental
data obtained with either type of  method are generally
characterized by a strong hysteresis phenomenon, the precise
nature of which is still a matter of debate\cite{Kierlik}.

 Recently, a new field of application for porous materials,
 devoted to the storage
or dissipation of mechanical energy,  has begun to develop
\cite{eros01}. This application is based on forced
intrusion-extrusion cycles of water in hydrophobic (non-wetting)
porous media. The range of pore sizes necessary for this
application is typically less than $10$nm, i.e. in the range of
mesopores as defined by IUPAC, and the energetic characteristics
of devices based on this process are directly related to the
hysteresis of the intrusion/extrusion cycles.  It is therefore of
importance to develop a quantitative understanding of hysteresis
phenomena involved in the condensation/drying transition in
mesoporous materials.

\smallskip

One promising material in this area are materials of the 'MCM41'
type \cite{Beck92}, in which the pore are essentially independent,
parallel  cylinders with diameters in the nanometer range. The
relative simplicity of these materials, in which the  pore
geometry is well understood and the connectivity between pores is
believed to be absent, makes them ideal for studying the
hysteretic behaviour. In the range of size of mesopores,  two
types of effects compete to induce  hysteresis: on one hand
kinetic effects associated with the phase transition control the
apparition of one phase when the material is saturated with the
other phase ; on the other hand effects related to the complexity
of the solid matrix determine the propagation of liquid/vapor
meniscii in the material. The contribution of those two types of
mechanism is not well understood. Disorder effects are often
addressed using a mean field approach to model capillary
condensation or drying. In these treatments, hysteresis is related
either to the disorder induced by the porous matrix (treated in
the simplest approaches by introducing different advancing and
receding contact angles) or to percolation effects ('pore
blocking' models). In such approaches quantitative predictions are
limited by the need of using a precise description of pore
geometry. As far as the kinetics of phase transition is concerned,
two types models have been proposed to deal with capillary
evaporation (also called as the drying transition)  in hydrophobic
systems.  The first approach is to envision the process as driven
by the nucleation of a vapour bubble. In this scheme, drying is
topologically equivalent to capillary condensation by nucleation.
An exemple of this approach can be found in the work of Restagno
\emph{et al.} \cite{restagno}, who used a macroscopic approach to
predict the behavior of critical nuclei in two and three
dimensional slit pores,  and made a comparison with a time
dependent Landau-Ginzburg simulation of condensation in two
dimensions. Talanquer and Oxtoby \cite{talanquer} implemented a
density functional theory for the nucleation in slit pores and
improved the macroscopic model by incorporating a line tension.
Bolhuis and Chandler \cite{bolhuis} combined the transition path
sampling method with molecular dynamics and Monte Carlo
simulations to study the drying transition path in narrow pores.
The second type of  model is based on the idea that density
fluctuations lead to a spinodal type of instability for a liquid
film between parallel plates. Under this category, one may for
example cite the work by Lum \emph{et al.} \cite{lum, lum2}, who
used Glauber dynamics Monte Carlo simulations for a lattice gas
confined in a slit pore with strongly hydrophobic walls. Their
work showed that the drying transition can be, in this situation,
driven by a large wavelength fluctuation of the density at the
interface.   Wallqvist \emph{et al.} \cite{wallqvist} also
considered the influence of attractive Van Der Waals forces on the
density fluctuations near the interface, and showed that these
interactions could strongly reduce the width of the interfacial
region, therefore reducing the fluctuations that lead to spinodal
decomposition.

\smallskip

In view of this rather confusing situation, it appears useful to
investigate the drying transition in well characterized materials,
and to attempt a quantitative comparison between model
calculations and model experiments in order to understand which route
towards drying can account for the experimental results. The work
presented in this paper represents a first step in this direction.

We first present an experimental study of intrusion-extrusion
cycles of water in hydrophobic MCM41. In these model materials
made of independent pores of cylindrical shape, effects related to
the disorder of the solid matrix such as "pores blocking" are not
expected to be important. The hysteretic behaviour of
intrusion/extrusion cycles should be a kinetic phenomenon
associated with the dynamics of the phase transition in the
confined system. We find that  the intrusion pressure of water is
governed by the Laplace law of capillarity and scales as the
inverse of the pore radius down to a pore size of $1,6 nm$. In
contrast,  the extrusion pressure is governed by the nucleation of
a vapor phase in the pores and varies much more rapidly with the
pore size than the intrusion pressure.

Part III addresses the nucleation of a bubble in a cylindrical
pore. We use a  simple macroscopic model based on classical
capillarity for calculating the energy barrier. The macroscopic
approach, in spite of its limitations when dealing with nanometer
sized pores, has proven to be quite robust down to very small
length scales and is well adapted here since it describes
successfully the intrusion process. We show that depending on the
ratio between the pore size and the Kelvin's radius, the shape of
the critical nucleus is either an annular cylindrical bump, or an
asymmetric bubble growing on one side of the cylinder.

Part IV is devoted to a quantitative comparison between  theory
and experiment. The plain classical capillarity model is in good
qualitative agreement with the data and describes quantitatively
well the temperature dependence of the hysteresis cycle. However
it fails to describe accurately the variation of the extrusion
pressure with the pore size. We show that excellent quantitative
agreement is recovered if one takes into account line tension
effects, i.e. the  energy of the three-phase line of the critical
nucleus. Experimental extrusion occurs when the energy barrier has
a constant value of about $40 k_BT$ for all pore sizes and
temperature investigated.

\section{Experimental section}
\subsection{Parent materials}

\begin{figure}[h]
    \bc
    \includegraphics[width=15.0cm]{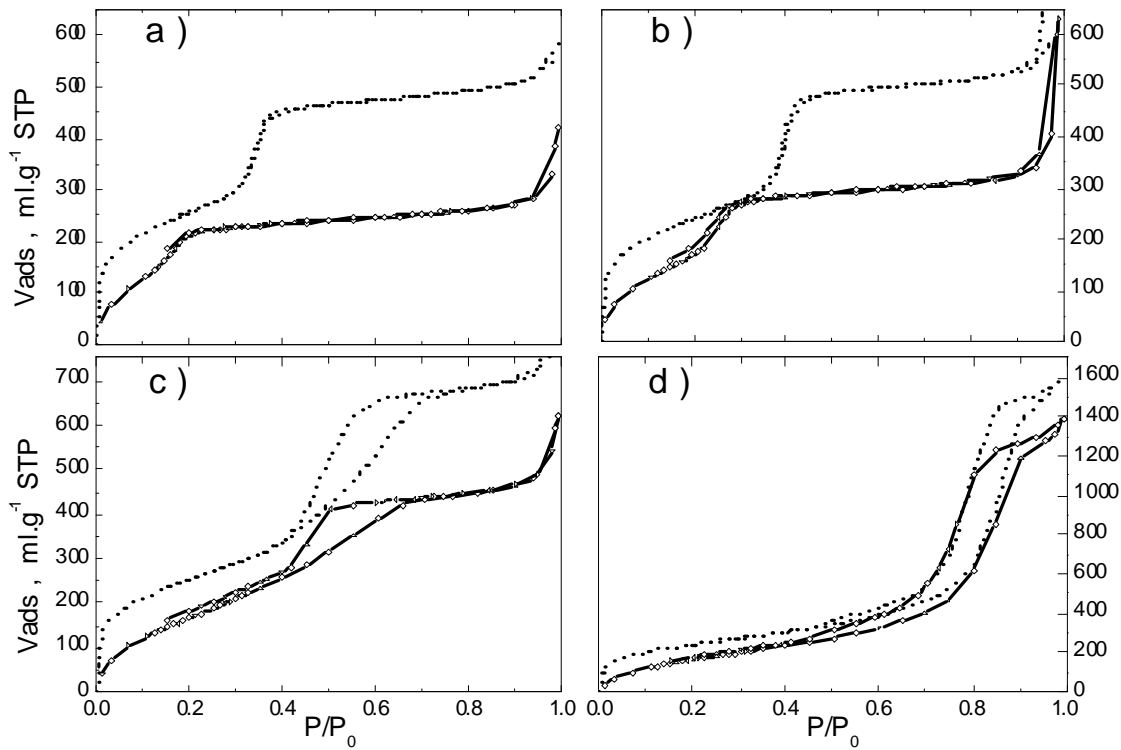}
    \caption{Nitrogen sorption isotherms of the parent
     silica supports (dotted line) and grafted corresponding
      materials (full line). Data are plotted per
       gram of bare silica in each case : a) MTS-1, b) MTS-2,
        c) MTS-3 and d) MTS-4. \label{figure1.fig}}
    \ec
\end{figure}

\begin{figure}[h]
    \bc
    \includegraphics[width=8.0cm]{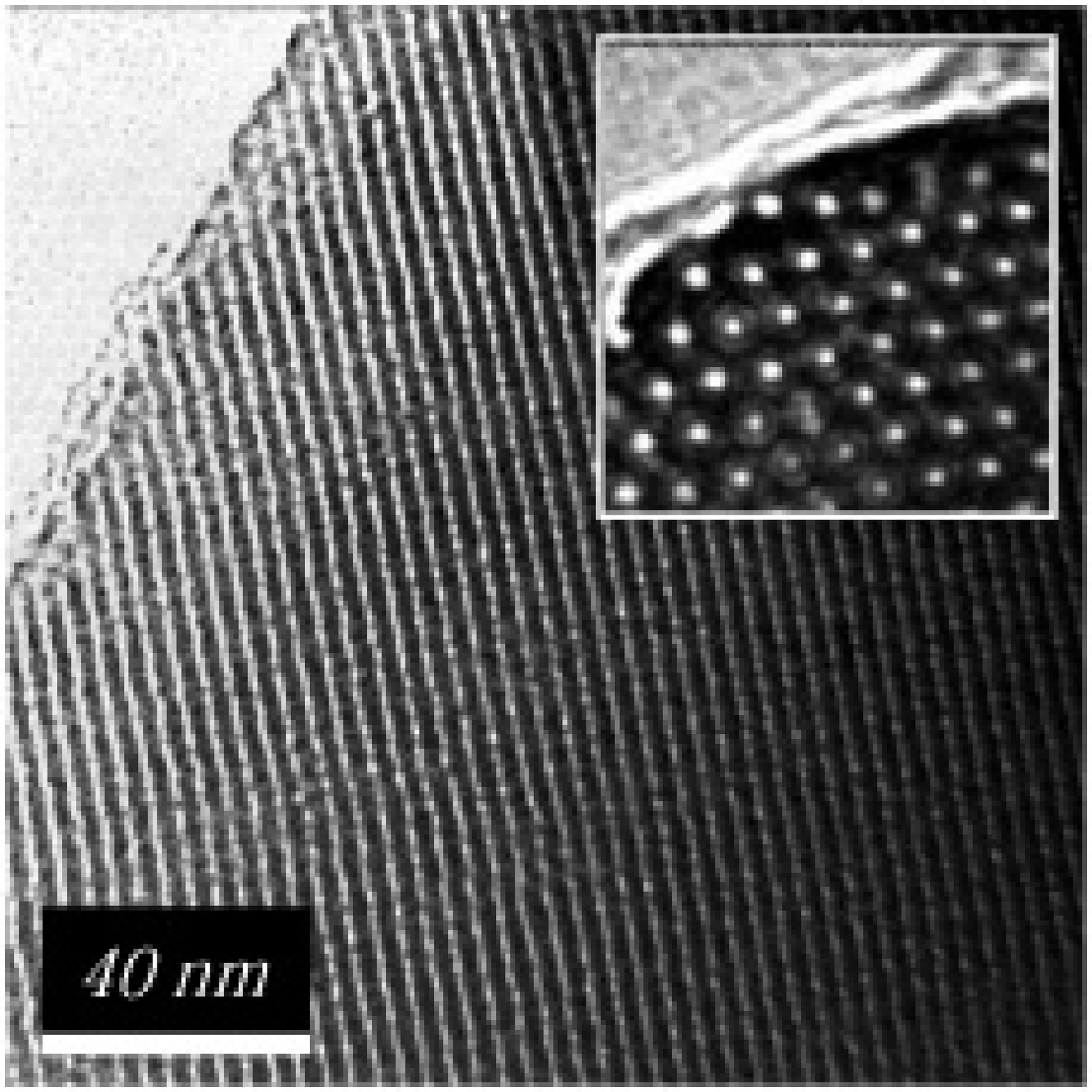}
    \caption{Transmission electron microscopy images on
    sample MTS-2. The hexagonal pore arrangement is clearly evidenced on the
     top view, when observed perpendicularly to the pore channels. \label{figure2.fig}}
    \ec
\end{figure}

The materials used in this work are micelle-templated silicas,
MTS, of the MCM-41 type synthetised from an alkaline silicate
solution in the presence of surfactants \cite{Beck92}:
hexadecyltrimethylammonium bromide (CTAB) and
octadecyltrimethylammonium bromide, were used in the synthesis of
samples MTS-1 and MTS-2 respectively. Following standard
procedures \cite{Beck92} \cite{Hanna02} , materials with larger
pores were obtained by incorporating a swelling agent of the
micelles such as trimethylbenzene (TMB). Samples MTS-3 and MTS-4
were prepared by incorporating TMB in the ratios TMB/CTAB of 2.7
and 13 respectively. Low temperature (77 K) nitrogen sorption
isotherms of the corresponding calcined materials are gathered in
Figure ~\ref{figure1.fig}. The sharp condensation steps for
samples MTS-1 and MTS-2 are typical from MCM-41 materials and
reveal low pore size distributions. The regularity of the
mesopores arrangement and the model structure are confirmed by
Transmission Electron Microscopy (TEM).  Images of sample MTS-2
are given in Figure ~\ref{figure2.fig} as an example. Less ordered
pore textures are expected when TMB is used as a swelling agent
(samples MTS-3 and MTS-4). Anyway, because the mechanisms involved
in the pore generation during the synthesis are similar for the
four samples, it is quite reasonable to believe that their porous
textures also consist in cylindrical and independent channels.
From nitrogen sorption data, the specific surface areas,
$S_{BET}^{p}$, were calculated using the BET theory
\cite{Brunauer38}. Mean pore radii, $R_{BdB}^{p}$, were calculated
from Broekhoff-De Boer theory (BdB) \cite{Broekhoff68} applied to
the relative pressure corresponding to the inflection point of the
desorption step. This determination was previously shown to be in
good agreement with nonlinear  density functional approaches
(NDFT) \cite{Ravikovitch00} and geometrical arguments for MCM-41
materials \cite{Galarneau99}. Mesoporous volume, V$_{P}^{p}$, were
measured as the amount adsorbed at the top of the condensation
step. Textural properties are gathered in table ~\ref{table1.tab}.

\begin{table}[h]
\bc
\begin{tabular}{|c|c|c|c|c|}
  \hline
  Parent materials             &   MTS-1   &   MTS-2   &    MTS-3   &     MTS-4
  \\ \hline
  $V_{P}^{p}$ / $ml.g^{-1}$    &    0.71   &    0.76   &    1.06    &     2.38
  \\
  $S_{BET}$ / $m^{2}.g^{-1}$   &     932   &    855    &    898     &      856
  \\
  $r_{BdB}^{p}$ / $nm$         &    1.8    &    2.0    &    2.4     &      5.9
  \\ \hline
  Grafted materials            &  MTS-1g   &   MTS-2g  &    MTS-3g  &     MTS-4g
  \\ \hline
  $n_{g}$ / $nm{-2}$           &    1.39   &    1.17   &    1.34    &     1.35
  \\
  $V_{P}^{g}$ / $ml.g^{-1}$    &    0.35   &    0.44   &    0.68    &     2.01
  \\
  $R_{BdB}^{g}$ / nm           &    1.3    &    1.5    &    2.3     &      5.6
  \\
  $R_{muff}^{g}$ / nm          &    1.3    &    1.6    &    2.0     &      5.4
  \\ \hline

\end{tabular}
\caption{Textural properties of the materials as determined from
nitrogen sorption experiments and carbon analysis.
\label{table1.tab}}  \ec
\end{table}


\subsection{Hydrophobized materials}
Hydrophobicity at the surface of the pores was generated by
covalent grafting of n-octyl-dimethylchlorosilane by a
pyridine-assisted reaction following a procedure described
elsewhere \cite{Martin01}\cite{proceedMRS02}. Grafted materials
are referred to as MTS-1g, MTS-2g, MTS-3g and MTS-4g. Grafted
chains densities, $n_{g}$, were determined from carbon analysis
and from $S_{BET}^{p}$ of the parent silica (see table
~\ref{table1.tab}). Nitrogen sorption isotherms on grafted-MTS
materials are reported per gram of bare silica in
figure~\ref{figure1.fig} together with sorption data on parent
materials. This correction allows a more intuitive visualization
of the textural modification induced by the grafting treatment
\cite{Rustamov01}. In a first approach, BdB theory can be applied
to estimate the mean pore radius of the grafted materials,
R$_{BdB}^{g}$ (see results in table ~\ref{table1.tab}), but the
presence of high amounts of organic ligands may strongly modify
the interactions between the pore surface and the adsorbate during
the sorption experiment and affect the desorption pressure.
Therefore, the values of pore radii determined by BdB theory on
these samples are questionable. To overcome this uncertainty, one
can estimate the pore radii of the modified samples by considering
the change in pore volumes measured by nitrogen sorption. A
similar approach was used by Fadeev et al \cite{Fadeev88} to
estimate the grafted layer thickness in similarly modified silica
gels. In the case of cylindrical pores, the volume which becomes
unaccessible to nitrogen molecules after the surface treatment is
actually represented as a muff according to this interpretation.
Using equation ~\ref{equation1.equ} \cite{Fadeev88}, the pore
radius for a modified sample, R$_{muff}^{g}$, can be calculated
(V$_{P}^{g}$ is the mesopore volume of the modified sample
expressed per gram of parent material).
 \be
    \label{equation1.equ}
    R_{muff}^{g}=R_{BdB}^{p}\sqrt{\frac{V_{P}^{g}}{V_{P}^{p}}}
 \ee
One of the interests of such a calculation in the present study is
that no assumption is done on the mechanisms of adsorption on a
grafted surface.

\subsection{Water intrusion-extrusion study}

\begin{figure}[h]
    \bc
    \includegraphics[width=8cm]{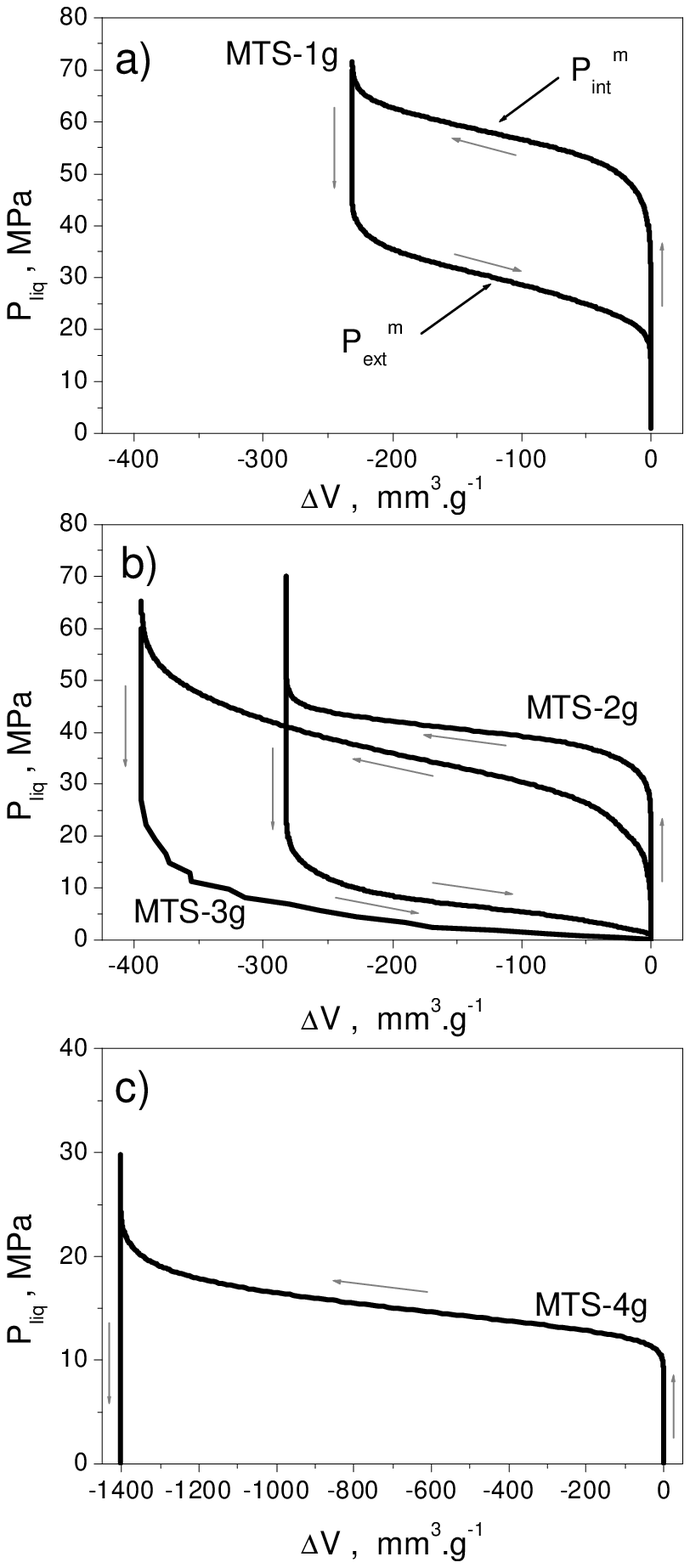}
    \caption{Intrusion/extrusion of water
    for the four grafted materials at 298 K : the pressure of
     the liquid $P_{liq}$ is reported as a function of
      the volume variation $\Delta V$ expressed per gram of bare silica. \label{figure3.fig}}
    \ec
\end{figure}

Water intrusion experiments were performed on a specially designed
apparatus described elsewhere \cite{ProceedSILICA2001}. About 2
grams of outgassed hydrophobized material were gathered with a
large excess of deionized water (compared to the corresponding
disposable pore volume) into a thermosealed shrinkable polymer
container. In a typical experiment, the pressure was first
continuously increased from atmospheric pressure, $P_{atm}$ to 80
MPa by means of a mechanical increasing constraint, and then
decreased back to $P_{atm}$. The time required for a complete
cycle was about 4 minutes. No significant difference was observed
in our experiments for 1 minute to 1 hour long cycles. The
pressure of the liquid and the volume variations were
simultaneously recorded. The volume values were corrected to
eliminate deformation and the compressibility contributions,
following a procedure described elsewhere \cite{Lefevre02}. The
representation of the water pressure as a function of the
corrected volume variation (expressed per gram of parent material)
are denoted P/V curves. In these plots, volume variations reflect
the cumulative intruded and extruded volumes of water into the
pores during compression and decompression steps respectively.
Therefore, the corresponding plots will be denoted as "intrusion
branch" and "extrusion branch". Tens of cycles could be recorded
on samples MTS-1g, MTS-2g and MTS-3g, allowing the system to stay
at $P_{atm}$ for 10 minutes between each run. Slight evolutions
were observed from the first to the third cycle, probably
corresponding to irreversible intrusion in some parts of the
material (see \cite{Eroshenko96}). The next
cycles were completely reproducible. Data presented in this paper
correspond to these last cycles designed as "stable" cycles. For
sample MTS-4g only one intrusion could be recorded as no water
extrusion took place during the decompression step, and even after
several hours at $P_{atm}$. For this material, spontaneous
extrusion of water does not take place. The corresponding P/V
curves are reported in Figure ~\ref{figure3.fig}. The absolute
value of the pore volume can be deduced from the water capacity at
the end of intrusion branch. This value is systematically smaller
than the pore volume determined by nitrogen sorption. This may be
a consequence of the strong difference between a good wetting of
the grafted chains by nitrogen and a non-wetting behavior in the
case of water. The mean intrusion (extrusion) pressure
$P_{int}^{m}$ ($P_{ext}^{m}$) related to the inflexion point of
the intrusion (extrusion) branch are gathered in table
~\ref{table2.tab} for the four selected samples. The dependance of
$P_{int}^{m}$ and $P_{ext}^{m}$ with  mean pore radius and surface
chemistry will be discussed in the next sections.

\bigskip

\begin{table}[h]
\bc
\begin{tabular}{|c|c|c|c|c|}
  \hline
  Grafted materials            &  MTS-1g   &   MTS-2g  &    MTS-3g  &     MTS-4g
  \\ \hline
  $P_{int}^{m}$ / $MPa$        &    59.5   &    44.4   &    35.0    &      14.4
  \\
  $P_{ext}^{m}$ / $MPa$        &    30.8   &    6.2    &    2.5     &       -

  \\ \hline
\end{tabular}
\caption{Mean intrusion and extrusion pressures determined from
the P/V curves for stable cycles for samples MTS-1g, MTS-2g and
MTS-3g and for the first intrusion for sample MTS-4g.
\label{table2.tab}} \ec
\end{table}

\subsection{Variation of intrusion pressure with pore size}

The Laplace equation is generally used to described the pressure
drop trough a curved interface. Assuming that the liquid / vapor
interfaces are spherical caps and that the pores are cylindrical,
the equilibrium pressure drop trough a meniscus in a
pore of radius $R_{p}$ is : \\
 \be
    \label{equation2.equ}
    \Delta P = - 2 \frac{\gamma_{LV}}{R_{p}}\cos\theta_{eq}
 \ee
where $\gamma_{LV}$ is the interfacial tension of the liquid/vapor
interface and $\theta_{eq}$ the equilibrium contact angle. Then,
assuming that the pressure of the vapor is negligible, and that
during the intrusion process the triple line advances by adopting
an advancing contact angle $\theta_{a}$
($\theta_{a}\geq\theta_{eq}$) we can in theory express the applied
pressure $P_{liq}$ required for intrusion by means of local
parameters $\theta_{a}$ and $R_{p}$. This relation, used to derive
pore-size distributions from mercury injection experiments, is
known as the Laplace-Washburn equation:
 \be
    \label{equation3.equ}
    P_{liq} = - 2 \frac{\gamma_{LV}}{R_{p}}\cos\theta_{a}
 \ee
\begin{figure}[h]
    \bc
    \includegraphics[width=12.0cm]{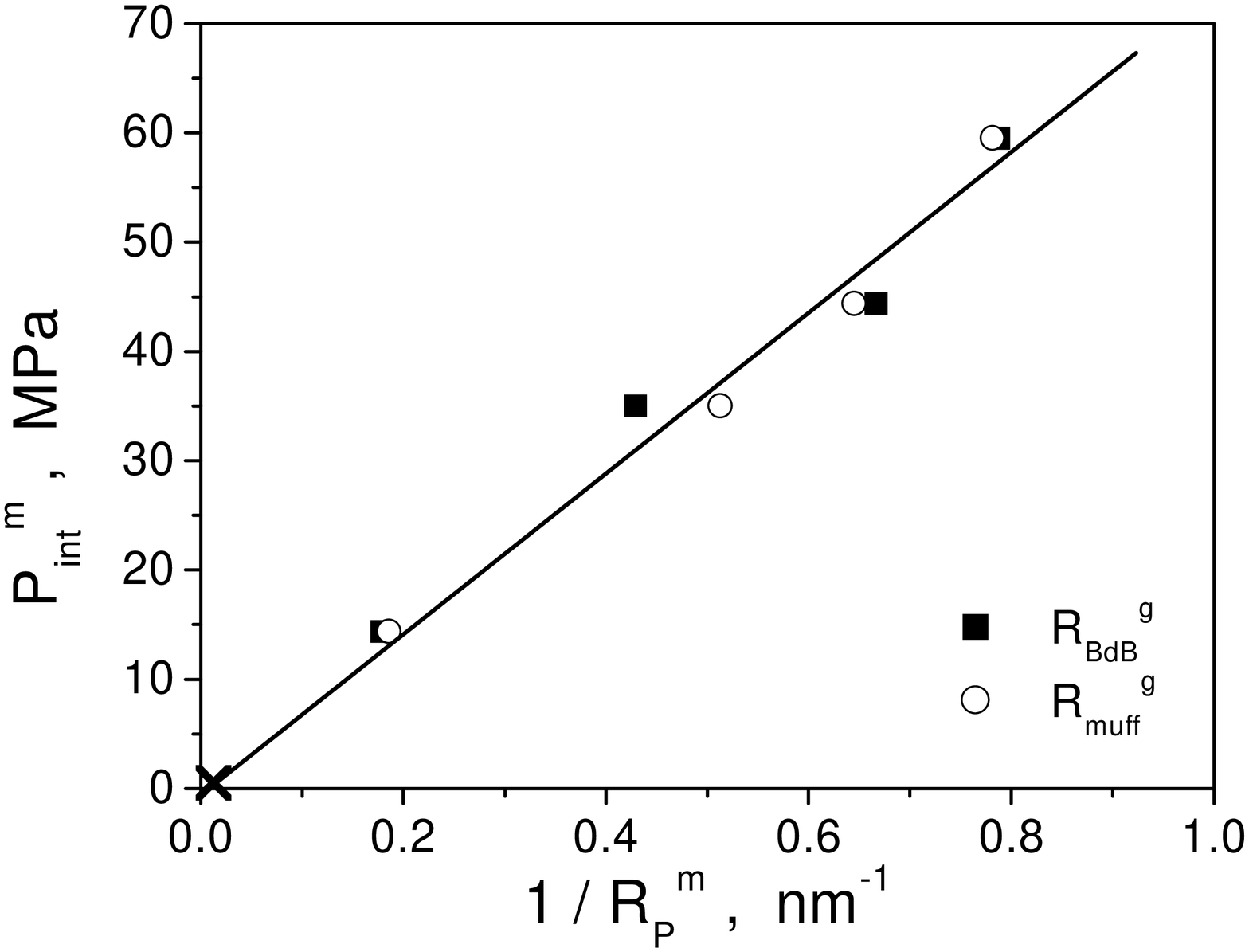}
    \caption{Mean intrusion pressure of water as a function of the inverse
     pore radius determined by nitrogen sorption for the four
      grafted materials.
      (The line is a guide for the eyes, the cross represents the origin).\label{figure4.fig}}
    \ec
\end{figure}
 If we now consider the set of data obtained on our materials, the
 validity of this equation to describe the variation of the intrusion pressure
 as a function of the pore size can be checked by plotting $P_{int}^{m}$
 against $R_{p}^{-1}$ (figure ~\ref{figure4.fig}). In this representation, a straight line is obtained
 with a good correlation when the pore radius is calculated
 following the muff model ($R_{muff}^{g}$). Several
 conclusions can be extracted from this result: (1) First, this
 linear plot includes with a reasonably good agreement the
 origin, which is consistent with the fact that $P_{int}\rightarrow 0$
 as $R_{P}\rightarrow \infty$. (2) Second, a  constant advancing
 contact angle for all the samples is expected, since the plot
 is linear. This indicates close values of
 the grafting densities $n_{g}$ for the four samples.
 Assuming the macroscopic value from the tables for
 $\gamma_{LV}$, the slope allows to estimate $\theta_{a}\approx
 120.3^o$, a reasonable value for this type of grafted materials.
  (3) Third, this results indicates that the
 representation of the grafted phase as a muff is suitable to
 describe the pore space and to estimate the pore size for this
 type of modified materials. (4) Finally, it is remarkable to note
 that the Laplace-Washburn equation is still successful to
 describe the intrusion of a non-wetting liquid into cylindrical mesopores of diameters as
 small as 2.6 nm, which had not been evidenced before for model
 materials (Eroshenko et al \cite{Fadeev96} \cite{Fadeev97mendel}
  \cite{Eroshenko95}. Gusev \cite{Gusev94} and Gomes et al \cite{Gomez00}
   reported interesting results for diameters
 down to 5 nm but in modified silica gels presenting a disordered porosity).
 Therefore, it could be said that confinement has no
 significant effect on the intrusion process -apart from the trivial effect expected
 from a macroscopic description.

\subsection{Variation of drying pressure with pore size}

\begin{figure}[h]
    \bc
    \includegraphics[width=12.0cm]{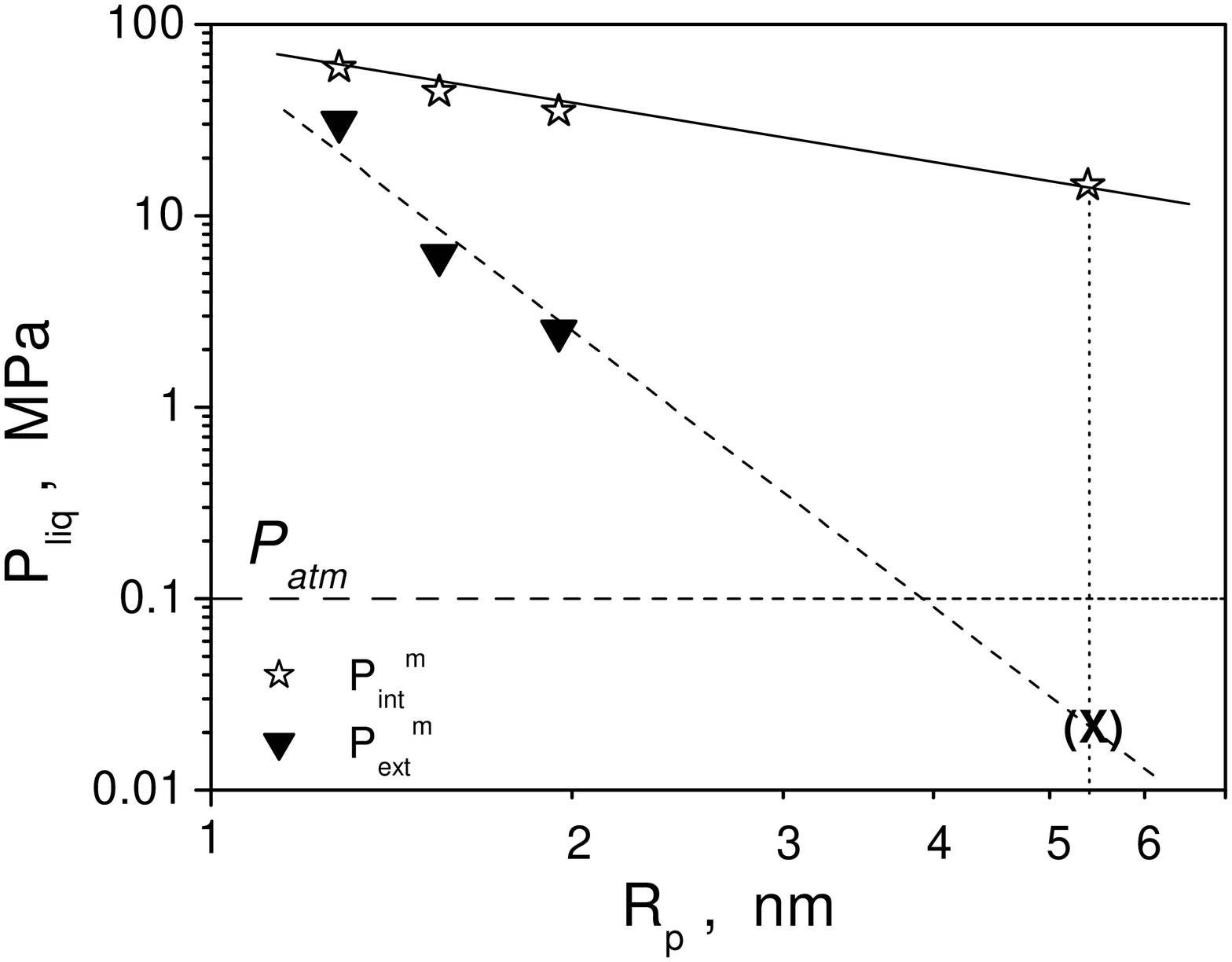}
    \caption{Mean intrusion and extrusion pressure of water as a function of
    the mean pore radius $R_{muff}^{g}$.\label{figure5.fig}}
    \ec
\end{figure}

We now consider the mean extrusion pressure, $P_{ext}^{m}$. The
conventional approach used to describe the withdrawal of mercury
is to consider a receding contact angle, $\theta_{r}$ to express
$P_{ext}$ by means of the Laplace-Washburn equation. As concluded
from the previous section, the surface properties (chemistry,
rugosity) for the present samples are believed to be quite similar
in terms of hydrophobicity as a single value of $\theta_{a}$ was
found for the four materials. Therefore, if the propagation of
menisci according to a receding contact angle was an accurate
description of the extrusion process, $P_{ext}$ should vary as
$R_{p}^{-1}$, in the same way as $P_{int}$. This dependency was
tested by plotting both $P_{int}^{m}$ and $P_{ext}^{m}$ as a
function of $R_{p}=R_{muff}^{g}$ in logarithmic scales in figure
~\ref{figure5.fig}. While the intrusion law (Laplace-Washburn)
appears as a line of slope $-1$, the tendency is completely
different for the extrusion law, as indicated by the experimental
mean extrusion pressures measured on samples MTS-1g, MTS-2g and
MTS-3g. $P_{ext}^{m}$ seems to be more sensitive to $R_{p}$ as
revealed by the larger slope  ($<-5$ compared to $-1$). In
addition, the extrusion pressure for the large pores material
(MTS-4g), which can be extrapolated from this tendency, is
expected to be lower than $P_{atm}$ (see the corresponding cross
between brackets on Figure ~\ref{figure5.fig}). This prediction is
in agreement with the fact that  this sample does not undergo
extrusion in the range of pressure accessible to our device. This
clearly demonstrates that the Laplace-Washburn equation is not
adequate to describe the pressure threshold for water withdrawal
in this set of model materials. Describing the extrusion process
as the propagation of a meniscus with a receding contact angle
would lead to pore size dependant values of $\theta_{r}$, which
are not consistent with the intrusion data on the same   samples.
To our knowledge, this failure of Laplace-Washburn equation had
not been clearly established before for mercury withdrawal,
probably because experimental results  have not  been reported for
the intrusion of mercury in model porous materials such as MCM-41.

Another mechanism has then to be considered to explain the
extrusion. Taking a pore full of liquid at a given pressure,
 the creation of the vapor phase has actually to be considered as the first
 step towards emptying, the second one being the propagation of
 the resulting  menisci.  In the range of pore size of our experiments,
 the propagation stage is clearly not the limiting one.
 We propose  that extrusion is governed by the nucleation of a vapor
phase in each pore independently, water withdrawal taking place
subsequently through  the fast,  non-equilibrium propagation of
menisci at the pressure for which the nucleation event has
occurred.

\subsection{Influence of temperature}

\begin{figure}[h]
    \bc
    \includegraphics[width=12.0cm]{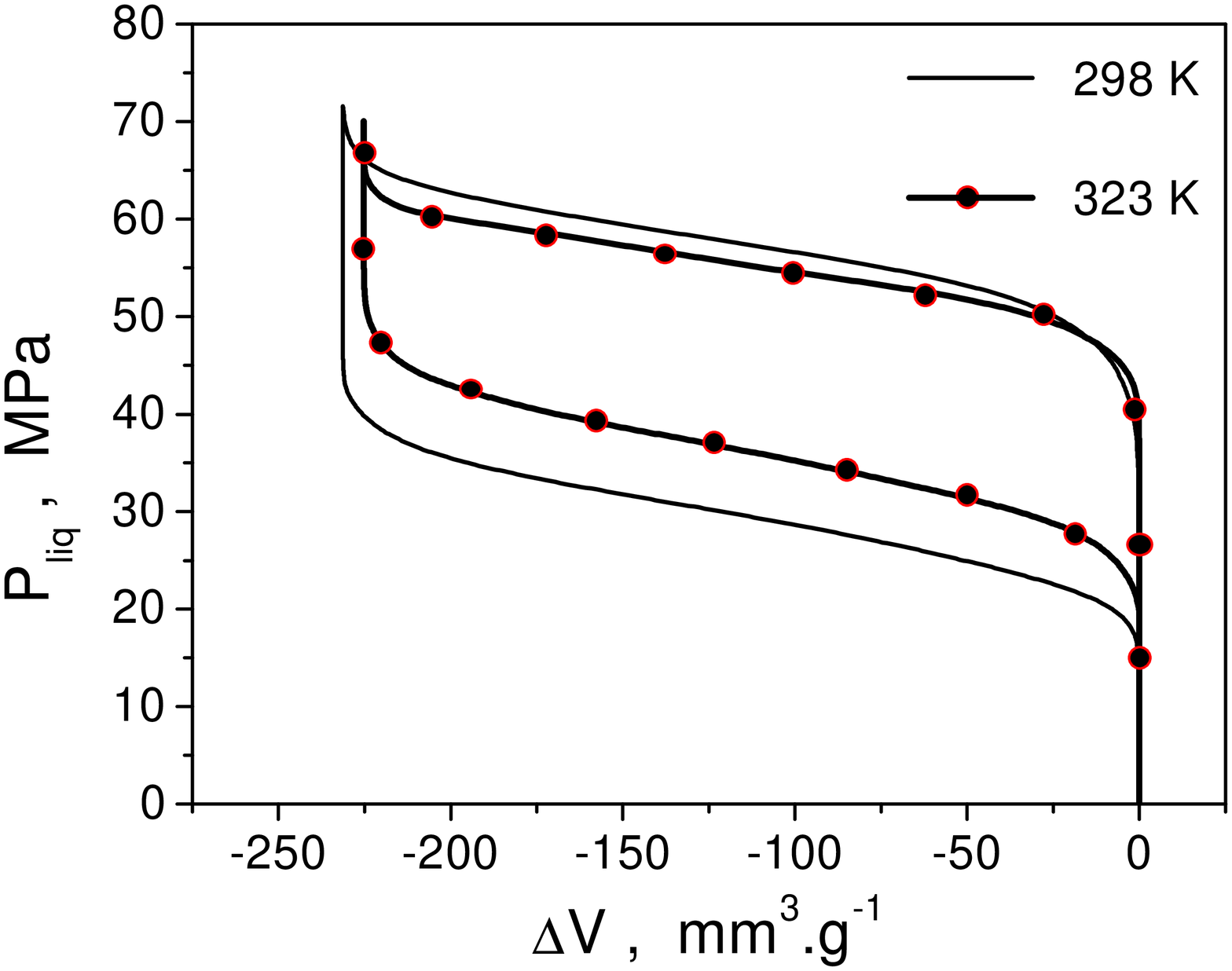}
    \caption{Comparison between the intrusion/extrusion cycles
     obtained at 298 K (line) and 323 K (line $+$ closed symbols) on
     sample MTS-1g.\label{figure7.fig}}
    \ec
\end{figure}

In phase transitions, nucleation processes are  thermally
activated and therefore highly temperature dependant.  A feature
arguing in favor of a nucleation mechanism for water extrusion in
our system is the behaviour of the intrusion-extrusion cycle when
the temperature is changed. Figure \ref{figure7.fig} shows a plot
of the cycle in one sample (MTS-1g), at two different temperatures
: $T_1=298$K and $T_2=323$K. The intrusion pressure at $323$K is
slightly lower than at ambiant temperature. This shift is
quantitatively consistent with the temperature variation of the
water surface tension. In contrast, the extrusion pressure
increases significantly with temperature : $P_{ext}= 30.8$ MPa at
298 K and $P_{ext}= 37.8$ at 323 K. This temperature shift of the
extrusion pressure is much more important than the one observed on
the intrusion pressure.
The high sensitivity of the extrusion pressure to temperature
confirms the hypothesis of a nucleation mechanism for this
process.

\medskip

\section{Modelling the nucleation path}
\label{theory}

In order to develop a quantitative understanding of the nucleation
process of a vapor phase in the hydrophobic MCM41, we calculate in
this section  the energy barrier to overcome for creating a
critical vapor nucleus in a hydrophobic cylinder. The words
"vapor" and "hydrophobic" are used here only for convenience ; the
calculation describes more generally the formation of a wetting
phase in a cylinder filled with the non-wetting phase. The only
limitation is that the contact angle must have a finite value :
the situation of perfect wetting with a wetting film adsorbed at
the wall is not described here. This calculation uses a relatively
simple macroscopic model based on classical capillarity. This is
justified for comparison with the experiments since the
macroscopic law of capillarity has been shown to describe
quantitatively the intrusion pressure.

Let us consider a pore of radius $R$. The interfacial tensions
between the solid boundary, vapour and liquid will be, as usual,
denoted by $\gsl$, $\gsv$ and $\glv$. The chemical potential of
the fluid molecules is  fixed at some imposed value $\mu$.
According to classical thermodynamics,  liquid-vapour coexistence
inside the pore is possible when the pressure difference between
the liquid and the solid,  $\Delta p = p_L-p_V$, verifies Kelvin's
equation \cite{Rowlinson} :
 \be
 \Delta p = 2(\gsl-\gsv)/R
 \label{eqn:kelvin}
 \ee
 Alternatively, $\gsl-\gsv$ can be replaced by
 $\glv \cos(\pi-\theta)$ where $\theta$ is the contact
angle of the  liquid  on the solid surface. For water on an
hydrophobic substrate,  $\theta> 90^o$, and the pressure inside a
vapour 'meniscus' is lower than in the surrounding fluid.

In the grand-canonical ensemble, a critical nucleus corresponds to
a saddle point of the grand potential. The grand potential of a
pore filled with liquid can be written as
 \be
\Omega_{L}=-p_L~V_{Pore}~+~\gsl~A_{SL} \ee
 while the potential of
a pore partially filled with vapor is
 \be
\Omega_{V}=-p_L~V_L~-~p_V~V_V~+~\gsl~A_{SL}~+~\gsv~A_{SV}~+~\glv~A_{LV}\ee.
 Here $V_L$ (resp. $V_V$) is   the volume of liquid (resp.
vapor) phase ($V_L+V_V=V_{Pore}$) and $A_{SL}$, $A_{SV}$, $A_{LV}$
are the  solid-liquid, solid-vapor and liquid-vapor surface areas.
With these notations,  the excess grand potential for a pore
containing a vapor nucleus can be expressed as
 \be
 \Delta\Omega=V_{V} \Delta p
+\gamma_{LV} A_{LV} + \gamma_{LV}\cos\theta ~A_{SV} \ee

In order to determine the shape of the critical nucleus, it will
prove convenient to introduce reduced quantities, by using $R$,
$R^2$ and $R^3$ as units of length, area and volume, respectively.
Using $\tilde{V}_V=\frac{V_V}{R^3}$,
$\tilde{A}_{LV}=\frac{A_{LV}}{R^2}$,
$\tilde{A}_{SV}=\frac{A_{SV}}{R^2}$, one obtains:
 \be \Delta\tilde{\Omega}~=~\frac{\Delta\Omega}{4\pi\glv
R^2}~=~2\delta \frac{\tilde{V}_V}{4\pi}
~+~\frac{\tilde{A}_{LV}}{4\pi}
~-~\cos(\pi-\theta)~\frac{\tilde{A}_{SV}}{4\pi} \label{eqn:barred}
\ee where $\delta=\frac{R}{2R_K}$ and $R_K=\frac{\glv}{\Delta p}$
is Kelvin's radius. Coexistence takes place when
 $\delta=\cos(\pi-\theta)$, and a
 a pore filled of liquid becomes metastable when  $\delta<\cos(\pi-\theta)$.

\subsection{Cylindrical critical nucleus : the annular bump}
\begin{figure}
\begin{center}
\includegraphics[height=8cm]{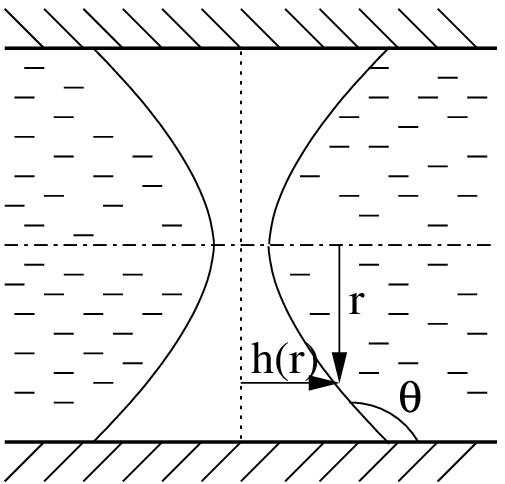}
\end{center}
\caption{Definition of the function $h(r)$ describing a nucleus
with cylindrical symmetry.} \label{fig:param}
\end{figure}

In view of the cylindrical symmetry of the problem, it is natural
to investigate first nucleation paths that preserve this symmetry.
The shape of a cylindrically symmetric vapour nucleus can be
described by a function $h(r)$, as shown in figure
\ref{fig:param}.
Using the   reduced variables $x=\frac{r}{R}$ and
$\psi(x)=\dfrac{h(r)}{R}$, the expression for the excess grand
potential can be written  as:
 \be
\Delta\tilde{\Omega}~=~2\delta \int\limits_{0}^{1}x \psi(x) dx
~+~\int\limits_{0}^{1} x
\sqrt{1+\left(\frac{d\psi}{dx}\right)^2}\,dx
-\cos(\pi-\theta)\psi(1) \ee

\begin{figure}
\begin{center}
\includegraphics[height=8cm]{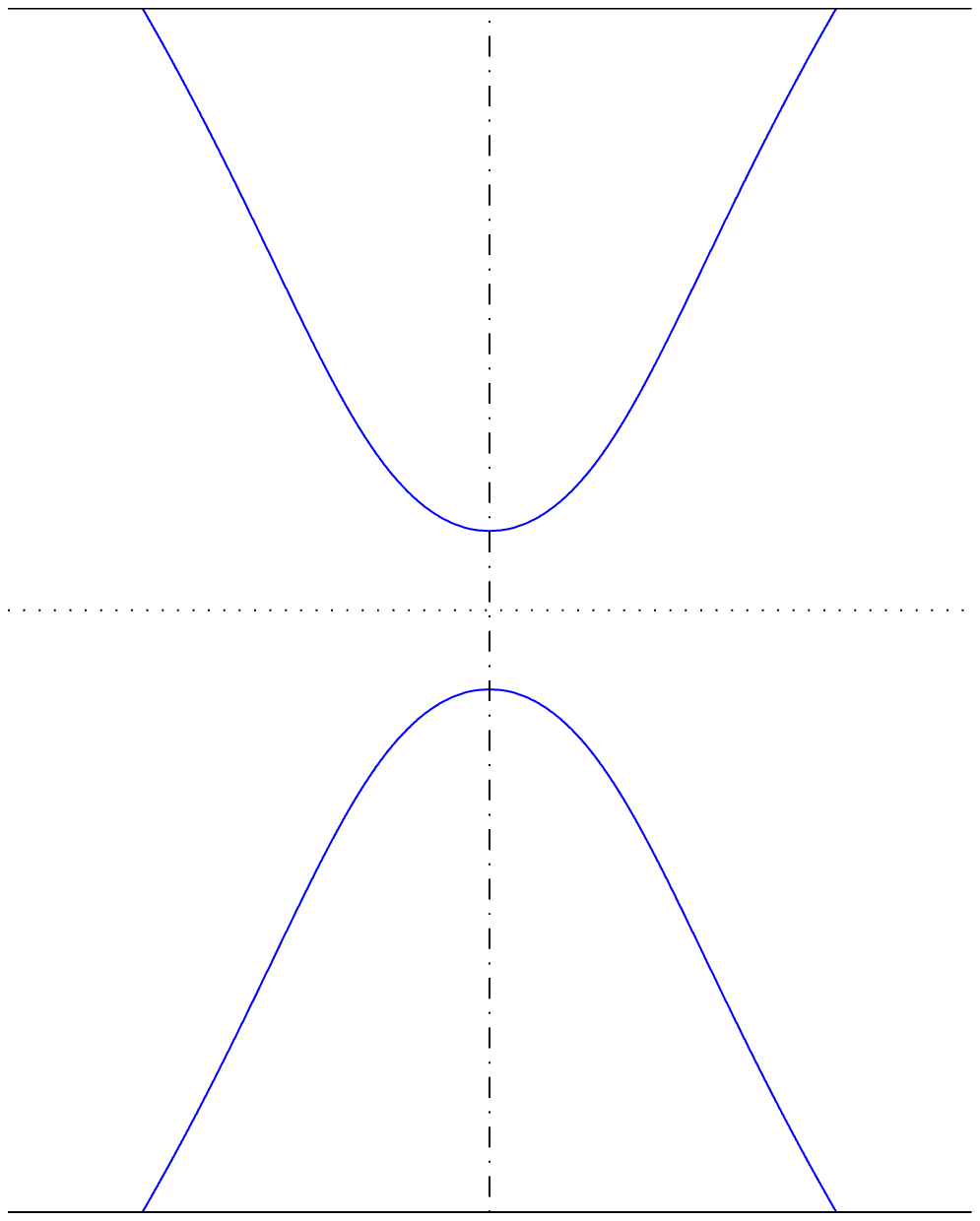}
\end{center}
\caption{Annular profile of the critical nucleus in a cylindrical
pore.$\theta=120^o$, $\delta=0.375$} \label{fig:profil}
\end{figure}

The shape of the nucleus is obtained by solving the Euler-Lagrange
equation $\frac{\delta \Delta\tilde{\Omega}}{\delta \psi(x)}$.
This gives the mechanical and contact equilibrium
 in terms of the local curvature and of the contact angle as
 \cite{kralchevsky}:
\begin{eqnarray}
\frac{1}{x}\frac{d}{dx}\left(\frac{x\frac{d\psi}{dx}}{\sqrt{1+(\frac{d\psi}{dx})^2}}\right)=2\delta
\label{eqn:derfonc:eqndiff} \\
\frac{\frac{d\psi}{dx}}{\sqrt{1+(\frac{d\psi}{dx})^2}}(x=1)=cos(\pi-\theta)
\label{eqn:derfonc:condlim}
\end{eqnarray}
Using the boundary condition (\ref{eqn:derfonc:condlim}),
the second order differential equation (\ref{eqn:derfonc:eqndiff})
 can be integrated once, which yields:
\be \frac{d\psi}{dx}=\frac{f(x)}{\sqrt{1-f(x)^2}}
~~~with~~~f(x)=\delta x+\frac{\cos(\pi-\theta)-\delta}{x} \ee At
$x_1=\frac{1-\sqrt{1-4\delta(\cos(\pi-\theta)-\delta)}}{2\delta}$,
$f(x_1)=1$ : the profile has a tangent parallel to the cylinder
axis, and the critical nucleus  displays an annular shape. The
profile can be explicitly written as
 \be
\psi(x)=\int\limits_{x_1}^{x}
\frac{\cos(\pi-\theta)-\delta(1-\rho^2)}{\sqrt{\rho^2-(\cos(\pi-\theta)
-\delta(1-\rho^2))^2}} \, d\rho \ee which can be integrated
numerically or by using special functions (cf. fig.
\ref{fig:profil}). The reduced energy barrier is obtained from
equation (\ref{eqn:barred}). Under equilibrium conditions, $f(x)=x
\times \delta$ and the meniscus is the classical spherical cap.

\subsection{Nucleation of a bubble in contact with the wall}
A second nucleation path is proposed without taking into account
cylindrical symmetry. It consists in the growing of a vapor bubble
on the cylindrical wall,  that eventually occupies the whole width
of the cylinder and forms  two spherical menisci.
 To
determine surface and volume energies along such a nucleation path
we have to determine the shape of the  asymmetric bubble as a
function of  volume. This shape is determined by a minimization of
the total energy, which implies mechanical equilibrium at each
point. This is a complex variational problem, that can only be
tackled numerically. We have used the \emph{Surface Evolver} code,
which provides an efficient way of solving this type of
variational problems. The resolution  is based on a discretisation
of the  surfaces using triangular facets \cite{surfaceevolver}.
Energies are obtained using  surface and line integrations.  Using
a gradient method, \emph{Surface Evolver} yields  the shape of
minimal surface energy for a given volume.

\begin{figure}
\begin{center}
\begin{tabular}{ccc}
\includegraphics[width=3.5cm]{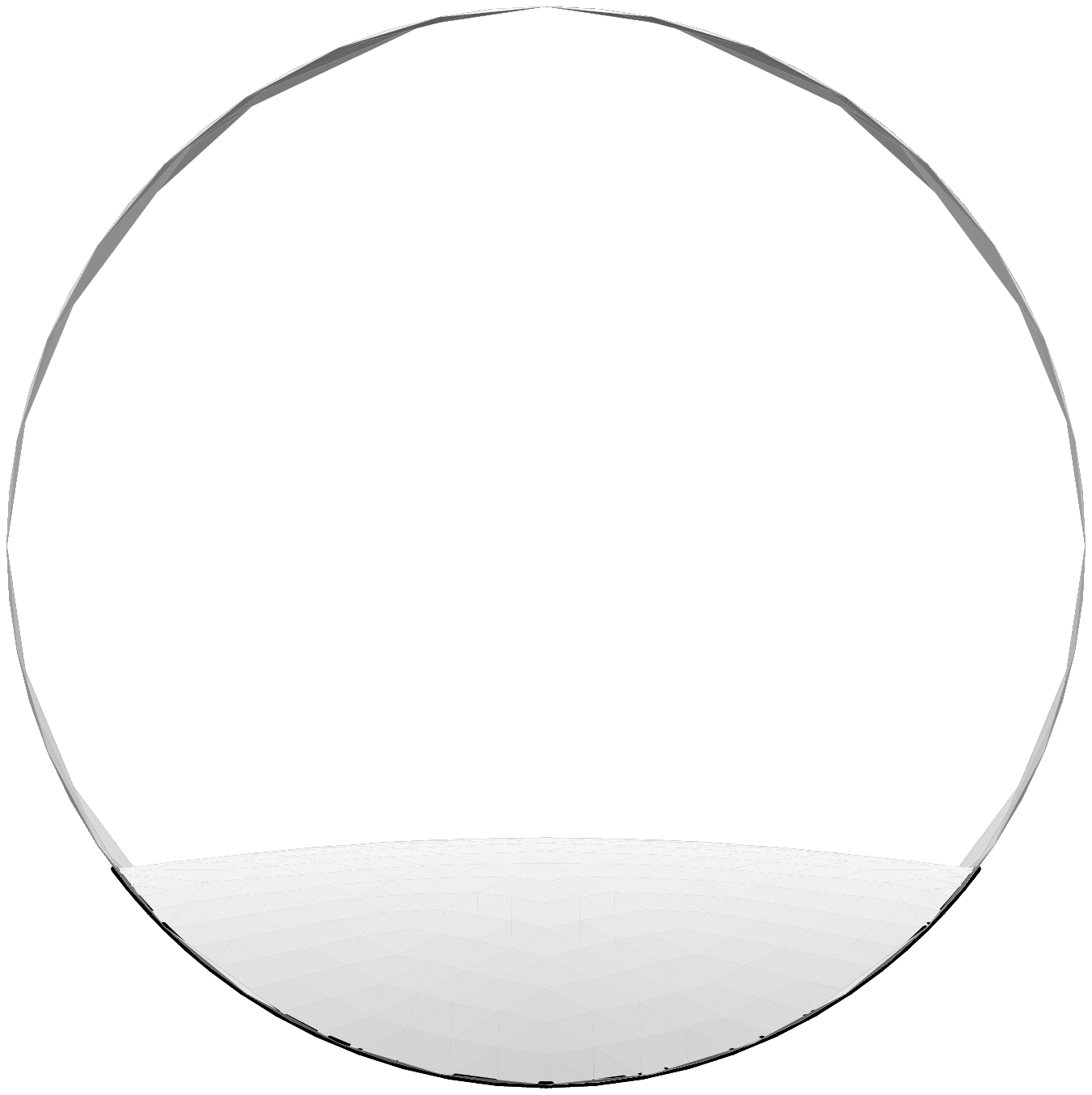} &
\includegraphics[width=3.5cm]{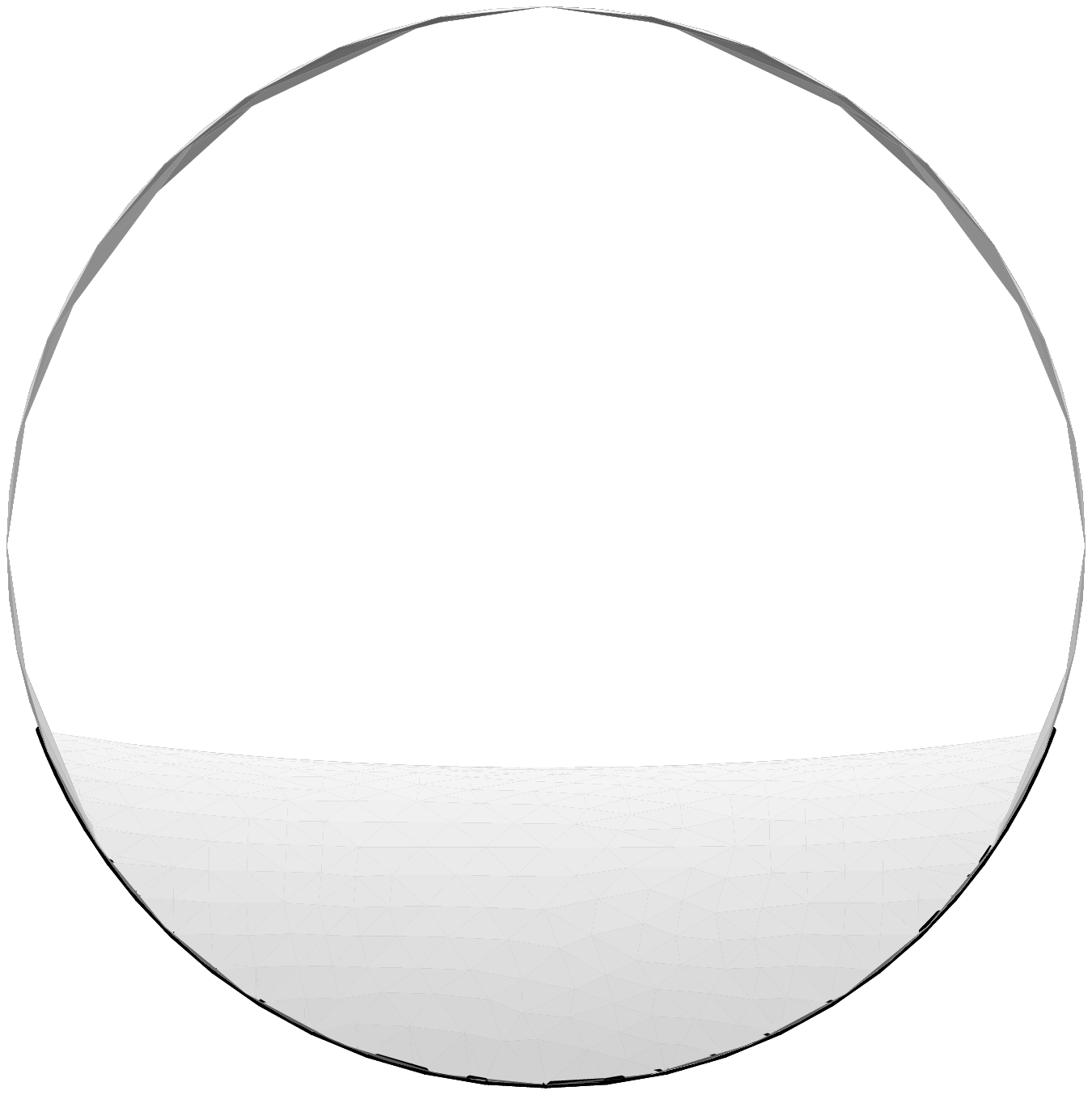} &
\includegraphics[width=3.5cm]{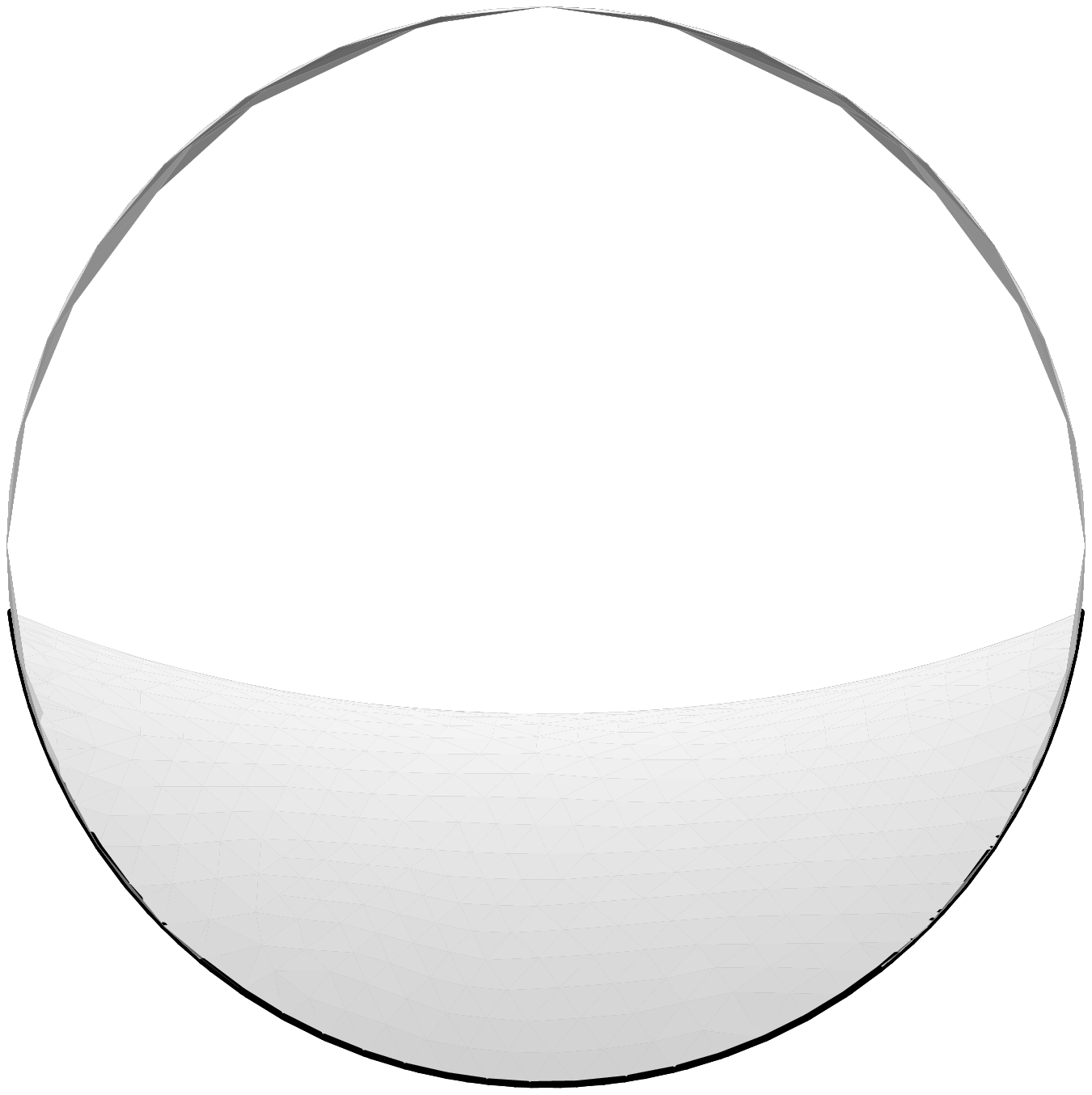} \\
\includegraphics[width=3.5cm]{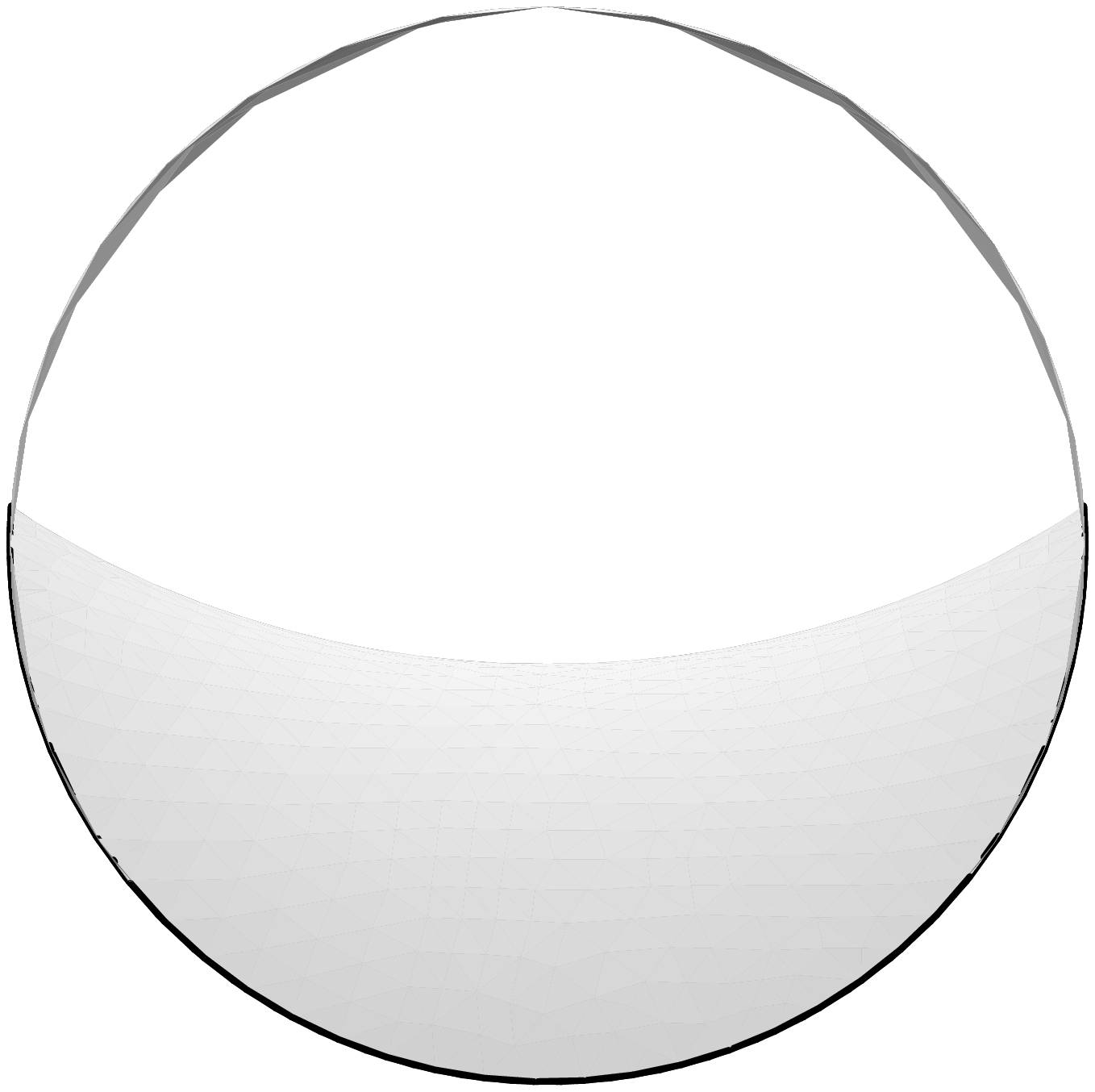} &
\includegraphics[width=3.5cm]{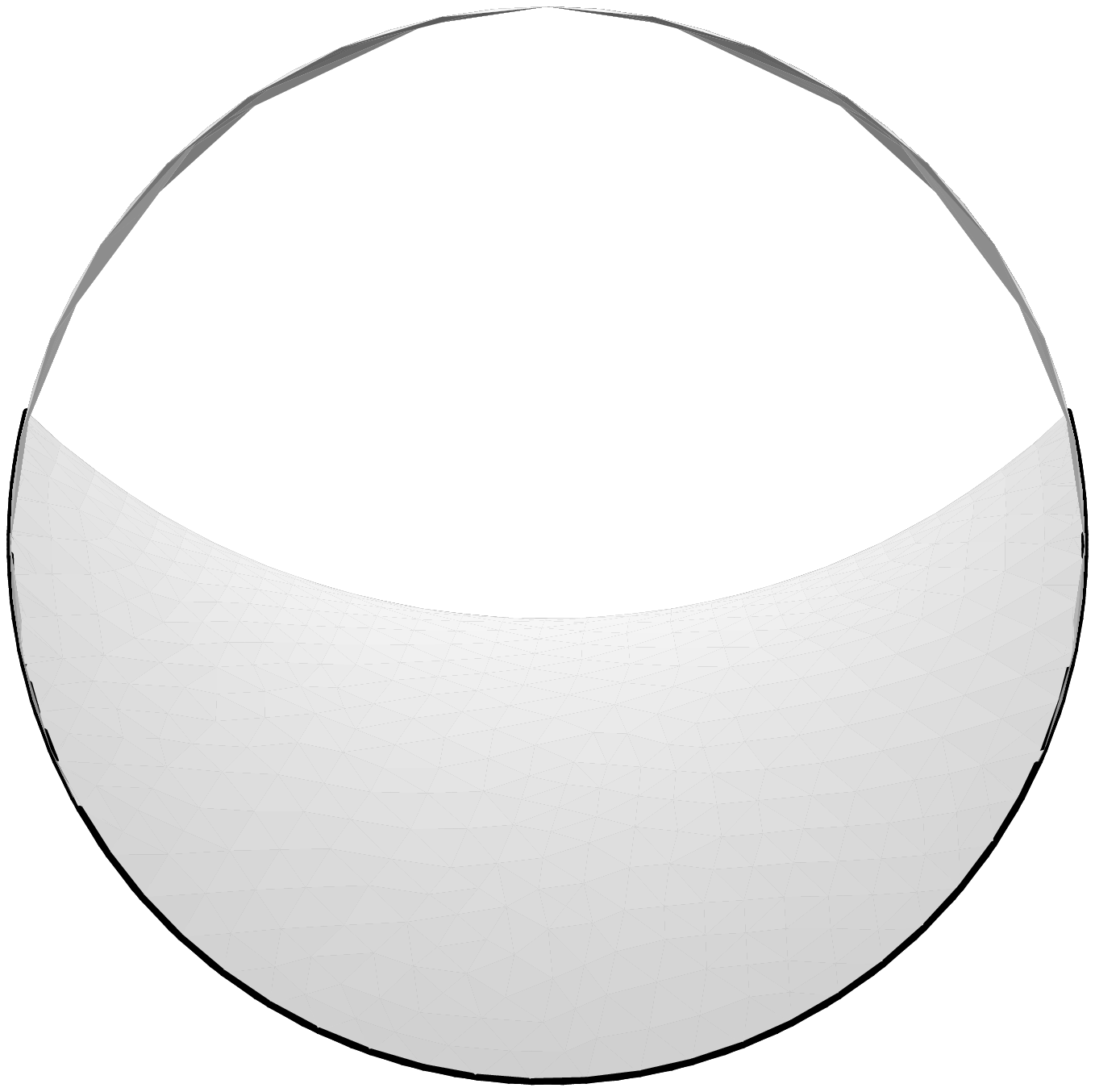} &
\includegraphics[width=3.5cm]{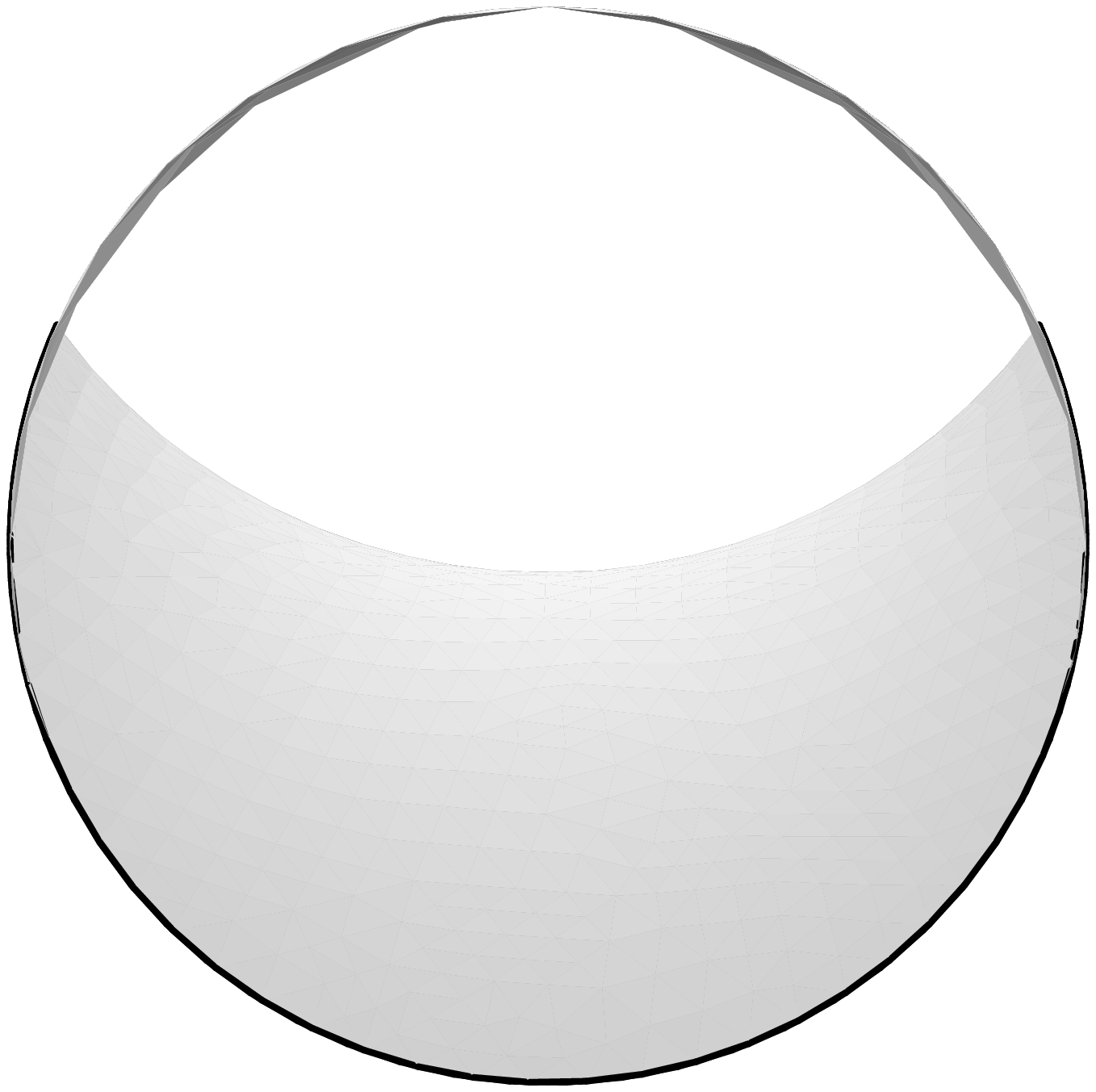}
\end{tabular}
\end{center}
\caption{Views of the vapour bubble growing on the surface of a
cylindrical pore,  for various values of the volume.
 Camera viewing axes is the cylinder axes. The contact angle is $120^o$. Only the
 liquid-vapour
  interface is displayed.}
\label{fig:volphilparoi}
\end{figure}

\begin{figure}
    \bc
    \begin{tabular}{ccc}
    \includegraphics[width=3.5cm]{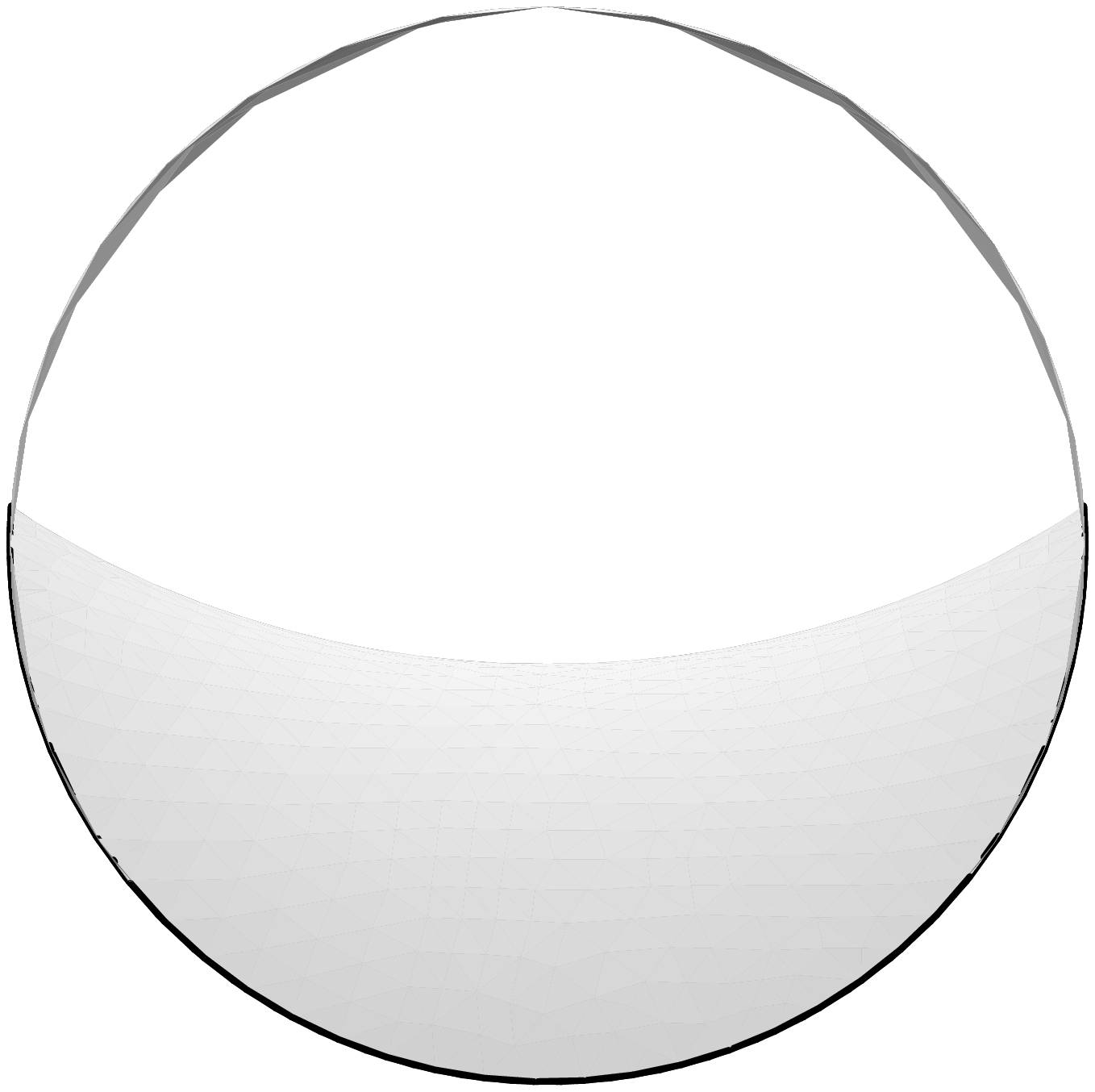} &
    \includegraphics[width=3.5cm]{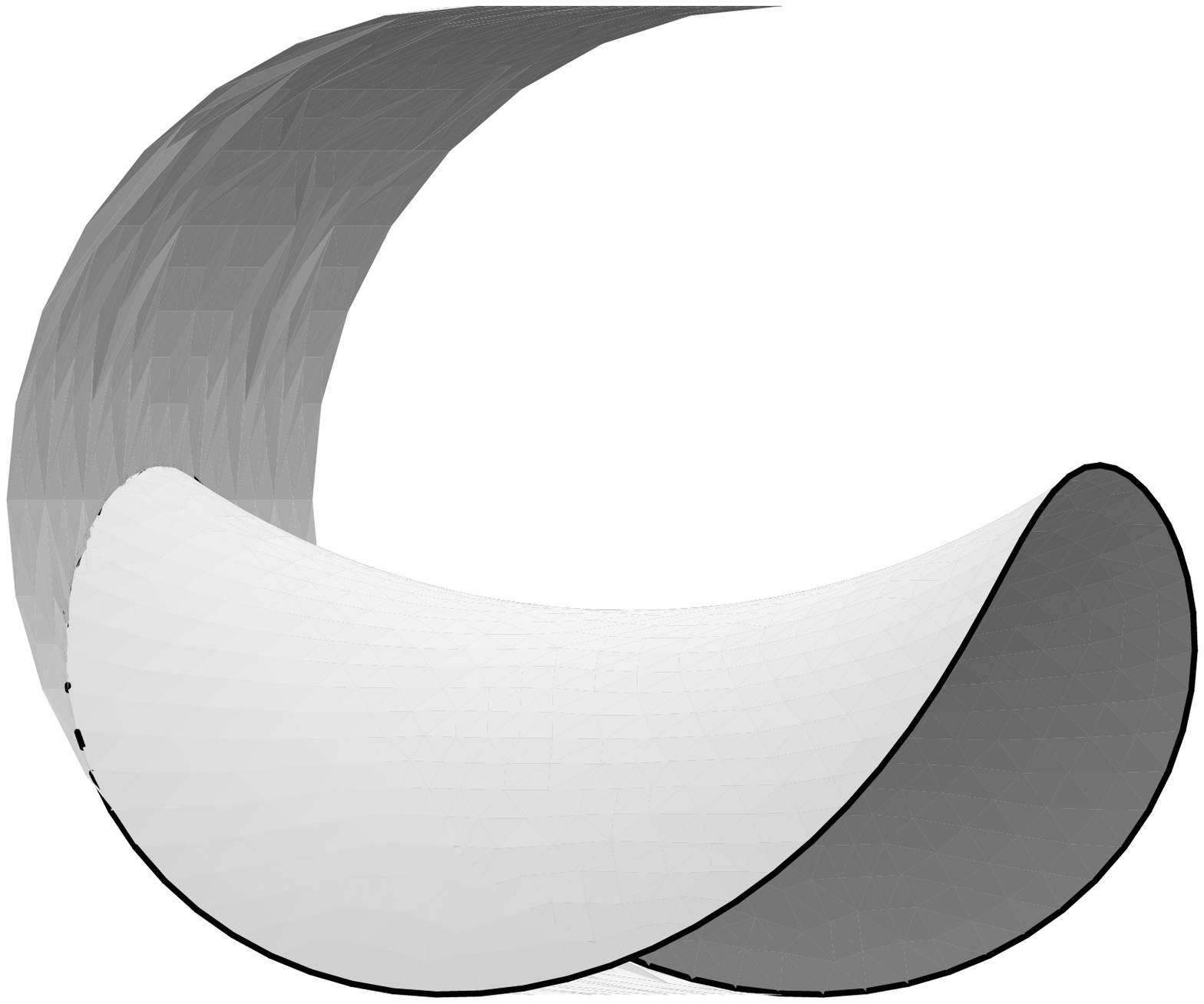} &
    \includegraphics[width=3.5cm]{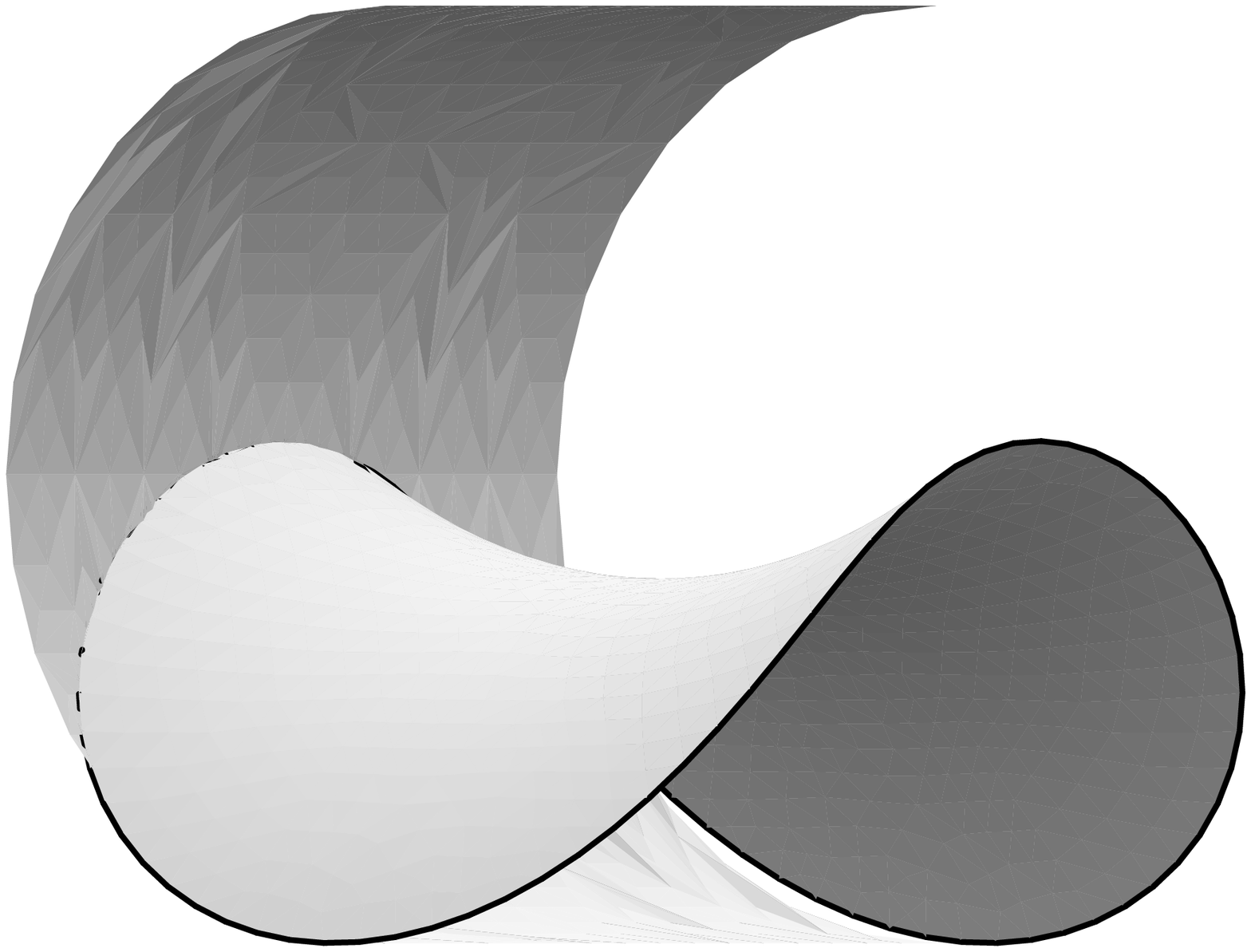} \\
    \includegraphics[width=3.5cm]{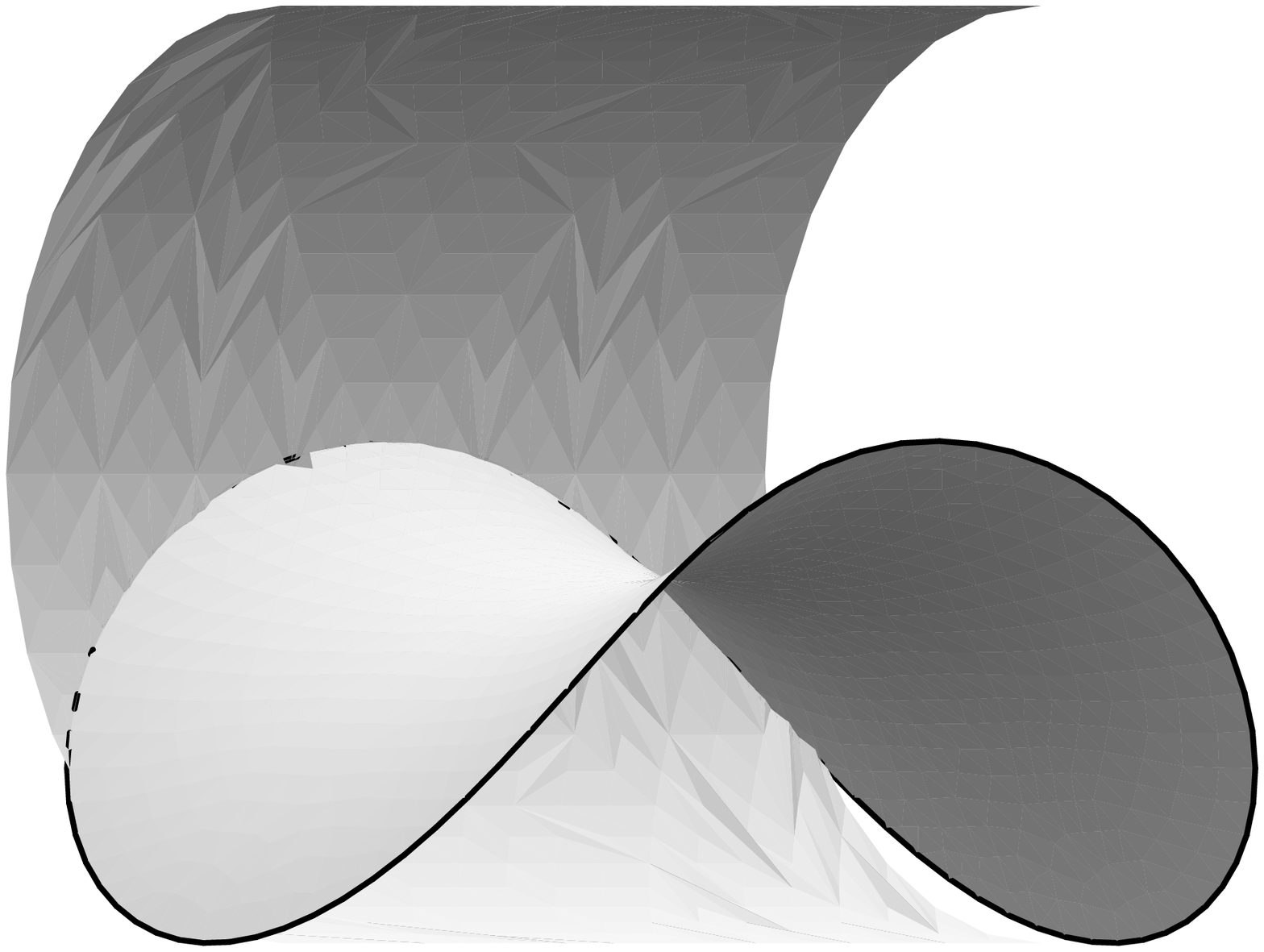} &
    \includegraphics[width=3.5cm]{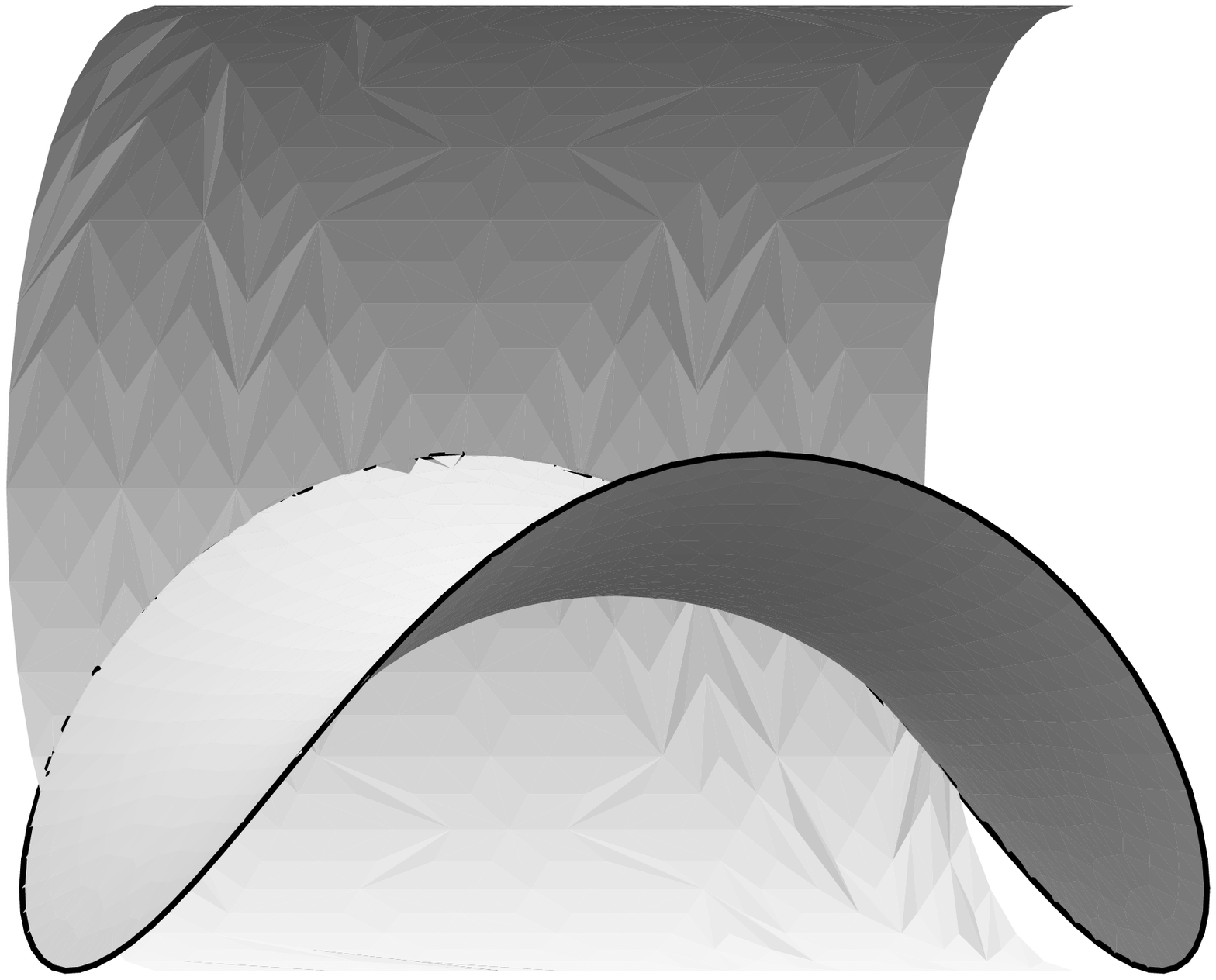} &
    \includegraphics[width=3.5cm]{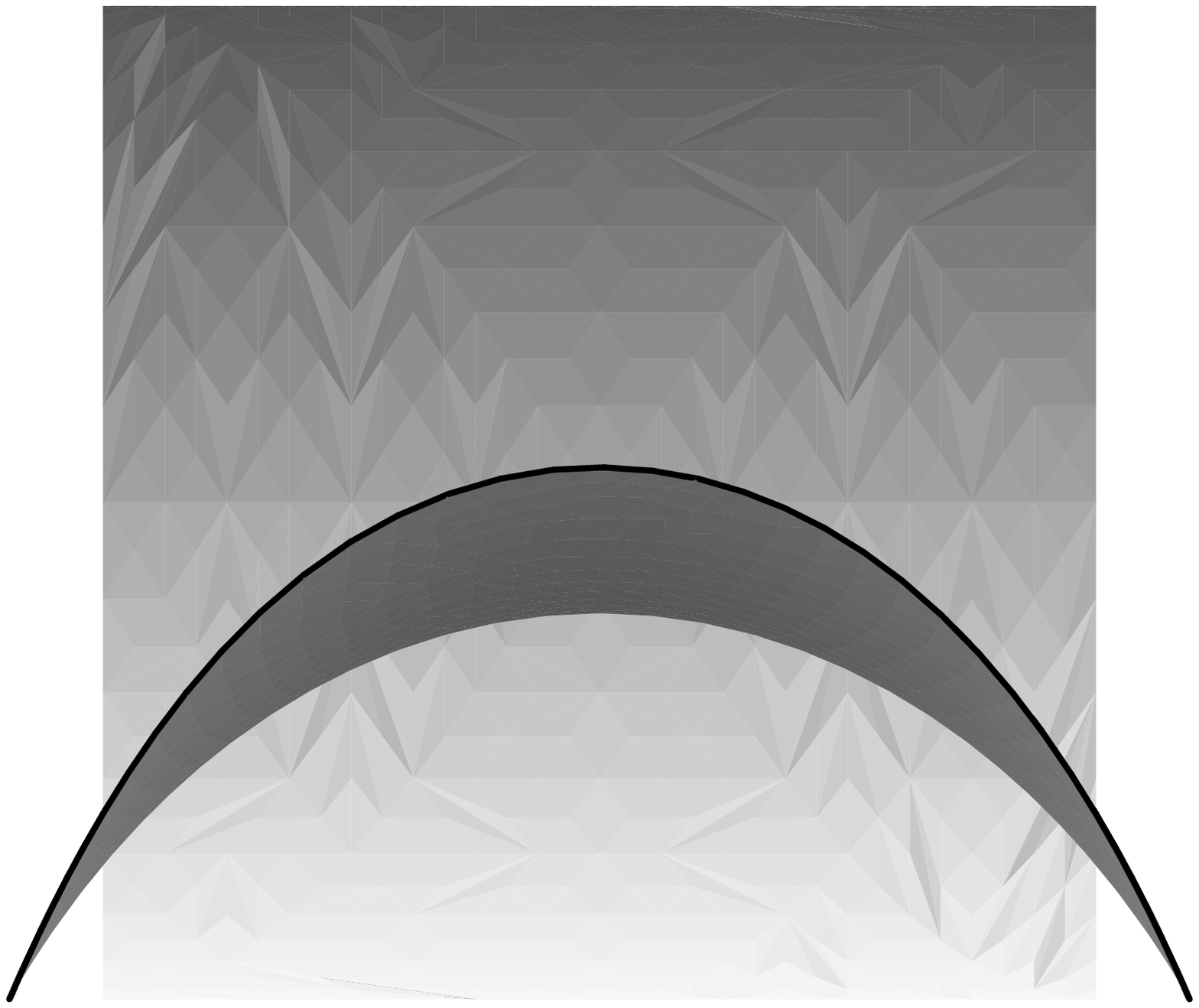}
    \end{tabular}
    \ec \caption{Several views of a vapor bubble in contact with the
    pore wall for an equilibrium contact angle fixed to $120¡$. Each picture is
    obtained from the previous one by an $18¡$ rotation to the
    left. Only one half of the cylinder has been displayed. The
    solid-vapor interface is not represented.} \label{fig:rotphilparoi}
\end{figure}

For a bubble in contact with the pore wall, the surface is made of
liquid vapor
 interface $S_{LV}$ of tension $\glv$ and solid vapor interface $S_{SL}$
 of tension $-\glv \cos(\theta)$ :
\be E~=~\int_{S_{LV}} \glv \,dA~+~\int_{S_{SL}} -\glv
cos(\theta)\,dA \ee For numerical stability, the solid-liquid
interface is removed from the code and surface energy  is
substituted by a line energy of the three phase line using the
Stokes theorem : \be E~=~\int_{S_{LV}} \glv \,dA~+~\int_{\partial
S_{SL}} -\glv \cos(\theta)~\vec{w_n}.\,\vec{ds} \ee with $\nabla
\times \vec{w_n}=\vec{F_n}$ and $\vec{F_n}$ being a  vector field
with zero divergence  ($\nabla \cdot \vec{F_n}=0$, that  reduces
to the normal vector for points on the solid liquid surface. In
the same spirit,  the volume integral is first reduced to  surface
integrals using the Ostrogradski theorem:
 \be
V~=~\int_{V}\,dV~=~\int_{S_{SL}\cup S_{LV}}
\frac{1}{3}\vec{r}.\vec{n}\,ds \ee
 and likewise replaced by a
contour integral as :
 \be V~=~\int_{V}\,dV~=~\int_{S_{LV}}
\frac{1}{3}\vec{r}.\vec{n}\,ds~+~\int_{\partial
S_{SL}}\frac{1}{3}\vec{w_r}.\,\vec{ds} \ee
 Results are presented
in figures \ref{fig:volphilparoi} and \ref{fig:rotphilparoi}. On
each picture, one can check that contact angle is equal to
$120^o$.

\begin{figure}
\begin{center}
\includegraphics[height=8cm]{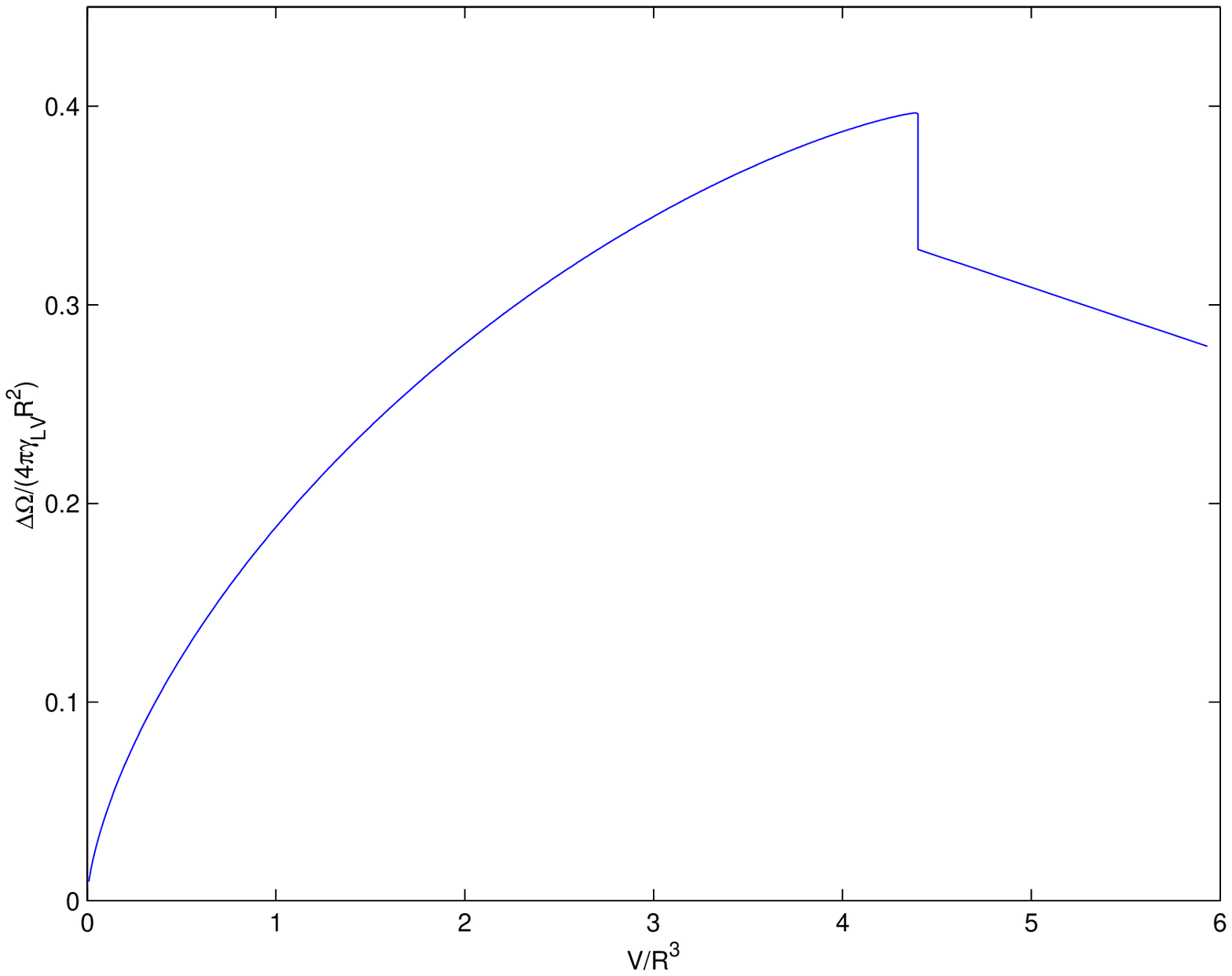}
\end{center}
\caption{Reduced excess grand potential of a vapor bubble in
contact with the pore wall as a fonction of the parameter angle.
$\theta=120^o$, $\delta=0.3$. For a maximum value of the volume,
the vapour bubble becomes unstable and degenerates into a
cylindrical tube. The energy barrier corresponds to this unstable
equilibrium shape.} \label{fig:nrjbull}
\end{figure}

 The reduced excess
grand potential, along the nucleation path obtained from equation
\ref{eqn:barred}, is plotted in figure \ref{fig:nrjbull}. For a
maximum value of the volume, the
 vapour bubble becomes unstable and a change of topology is observed,
 with formation of a vapour cylinder occupying the whole cylinder
 width and limited by two spherical caps.
 The energy barrier along this path is obtained for the maximum
 value of the bubble energy, which is reached at this stability
 limit. The critical nucleus  corresponds in this case to the
 unstable endpoint in a family of bubble like surfaces.

\begin{figure}
\begin{center}
\includegraphics[height=8cm]{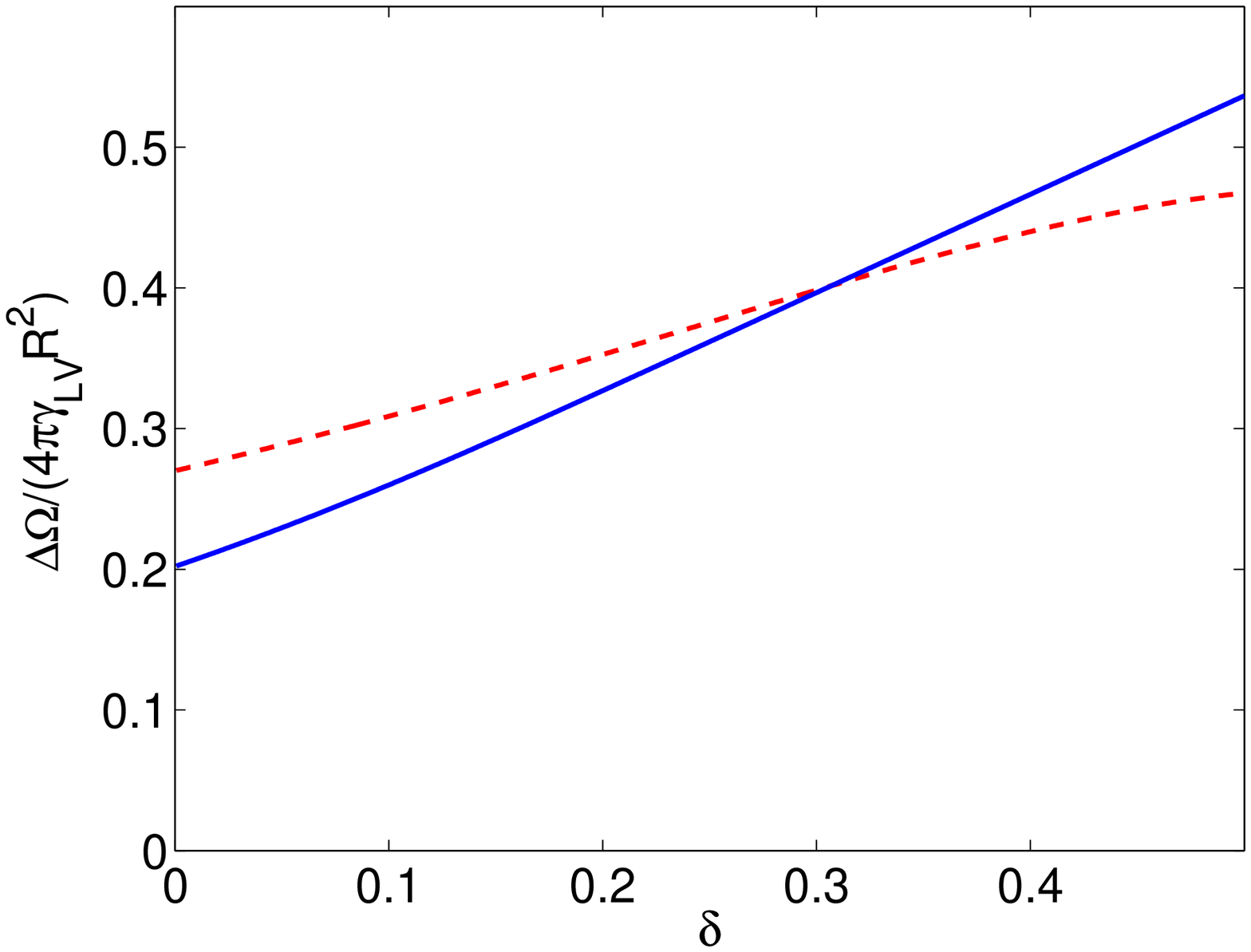}
\end{center}
\caption{Reduced energy barrier as a function of $\delta$, for
$\theta=120^o$. Full line: bubble configuration. Dashed line:
annular bump configuration} \label{fig:barredcomp}
\end{figure}

A comparison of the reduced energy barriers for the two nucleation
mechanisms is presented in figure \ref{fig:barredcomp}, for
$\theta=120^o$. Depending on the value of $\delta$, the asymmetric
nucleation path (low delta, i.e. high metastability) or the
annular nucleus (low metastability) is favoured.

\subsection{An approximate expression for the energy barrier of nucleation}
\begin{figure}
\begin{center}
\includegraphics{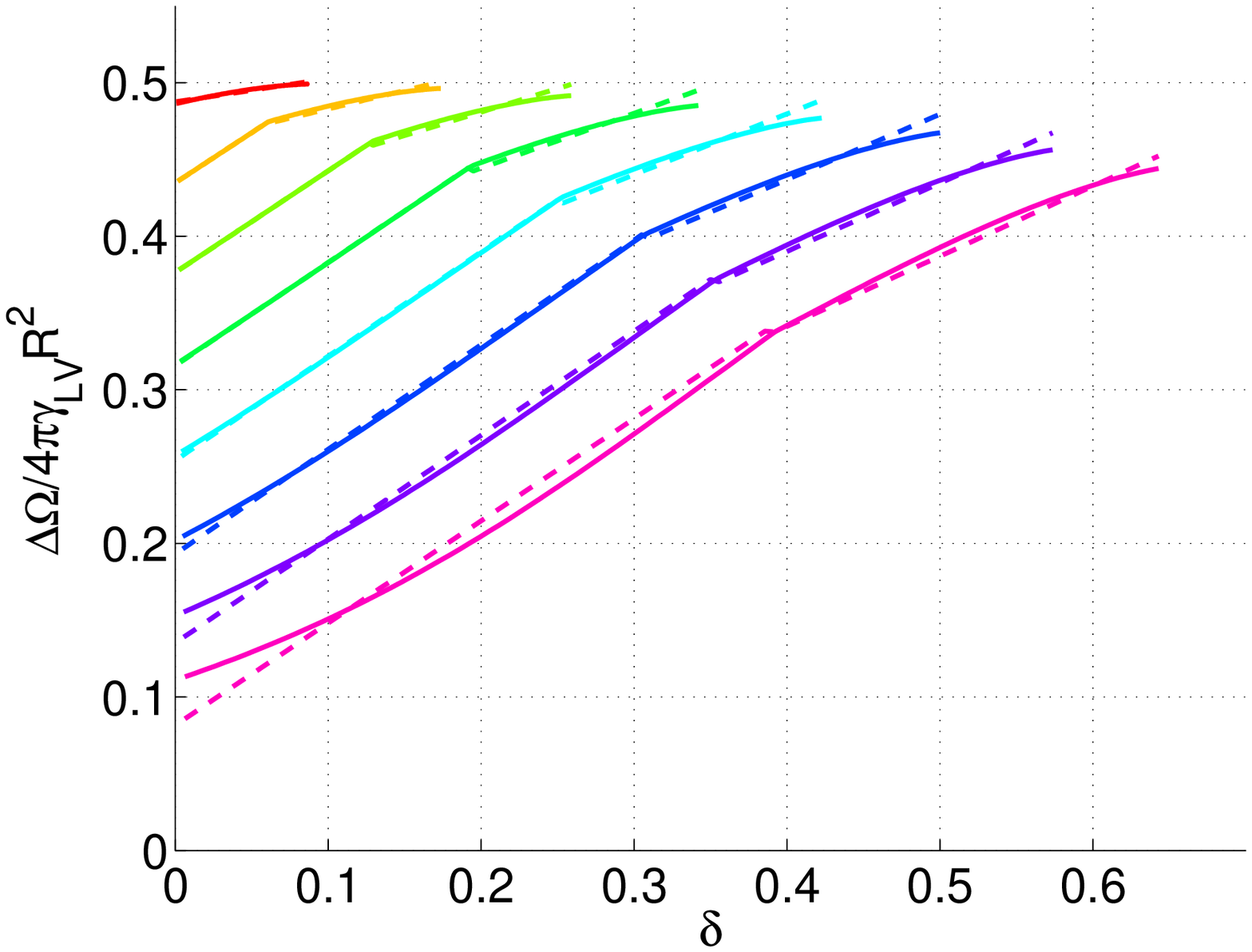}
\end{center}
\caption{Reduced energy barrier as a function of $\delta$ obtained
in the most favourable configuration (bubble or cylindrical bump).
From left to right, the values of the contact angle $\theta$ are
95$^o$, 100$^o$, 105$^o$, 110$^o$, 115$^o$, 120$^o$, 125$^o$,
130$^o$. The dashed lines are the best fit of each curve with a
piecewise linear function, as described in the text.}
\label{barredcomp}
\end{figure}

Figure  \ref{barredcomp} gathers the numerical results for the
energy barrier obtained for various values of the contact angle,
as a function of $\delta$. For each value of $\delta$, only the
energy barrier corresponding to the more favourable configuration
(bubble or annular bump) is represented.
A good approximation for estimating the reduced energy barrier is
to write it as  a piecewise linear function of the dimensionless
parameter $\delta=R\Delta p/2\glv$.  In this approximation,
$\Delta \Omega$ is written in the form
\begin{equation}
\Delta \Omega=\Delta p K_{1}(s) R^3+\glv K_2(s) R^2
\label{eqn:approx}
\end{equation}
where $K_1(s)$ and $K_{2}(s)$ are functions only  of the contact
angle and of the shape $s$ of the nucleus (bubble, $s=b$, or
cylindrical, $s=c$)
 and do not depend on the size of the capillary. The relative
error made by using approximation \ref{eqn:approx} instead of the
actual value of the energy barrier is at most $\pm 1\%$ for values
of $\theta$ ranging between $95^o$ and $130^o$. Numerical values
for $K_1(s)$ and $K_{2}(s)$  in this range of contact angle are
given in table \ref{tablex} for various contact angles.

It is of interest to notice that equation \ref{eqn:approx} if of
the general form that would be obtained if the critical nucleus
was keeping a constant shape when the radius $R$ of the capillary
is varied. It is easily seen from equation \ref{eqn:barred} that
if this were the case, the resulting energy barrier would be of
the form $\Delta \Omega = \gamma_{LV} R^2 f(R/R_K)$, with $f$ a
dimensionless function. Equation \ref{eqn:approx} is of this form,
$f$ being moreover closely approximated by a linear function.


\begin{table}[h]
\bc
\begin{tabular}{|c|c|c|c|c|c|c|}
  \hline
  $\theta$&$K_1(b)$& $K_1(c)$&$K_2(b)$&$K_2(c)$&$K_3(b)$&$K_3(c)$
  \\ \hline
   95.0000  & 4.06  &0.94 & 6.16& 6.13 & 11.85& 12.56\\
  100.0000  & 4.11  &1.51 & 5.46& 5.77 & 12.00& 12.56\\
  105.0000  & 4.17  &1.93 & 4.73& 5.27 & 12.16& 12.56\\
  110.0000  & 4.22  &2.24 & 3.97& 4.68 & 12.28& 12.56\\
  115.0000  & 4.27  &2.49 & 3.19& 4.04 & 12.38& 12.56\\
  120.0000  & 4.28  &2.67 & 2.42& 3.35 & 12.43& 12.56\\
  125.0000  & 4.25  &2.79 & 1.70& 2.67 & 12.46& 12.56\\
  130.0000  & 4.18  &2.87 & 1.02& 1.99 & 12.48& 12.56\\ \hline
\end{tabular}
 \caption{Dimensionless constants $K_1(s)$, $K_2(s)$ and
$K_3(s)$ for calculating approximately the nucleation barrier,
according to equations \ref{eqn:approx} or \ref{linetension}.
$s=c$ corresponds to the cylindrically symmetric nucleus, observed
at large $\delta$, while $s=b$ corresponds to the asymmetric
bubble (favoured at small $\delta$). \label{tablex}}  \ec
\end{table}

\section{Comparison between theoretical and experimental results}

In the usual theory of thermally activated nucleation, the number
$\bar n$ of critical nucleus created per unit time  and unit
volume (here unit length) in the system writes
\begin{equation}
 \bar n =(b\tau )^{-1}e^{-\Delta \Omega/k_BT}
 \label{nucleation}
\end{equation}
 The prefactor of this Arrhenius law includes
 a microscopic length $b$ and a microscopic time $\tau$
 whose value can depend on the pore size but in a much
 weaker way than the exponentiel factor. Hence the essential variation
 in the nucleation rate $\bar n$ comes from this exponential
 factor. Experimentally, the transition is observed when the
 probability of creation of a critical nucleus in an average capillary of
 length $ L$ over an experimental time $t$ becomes unity, i.e. $\bar n Lt \propto 1$.
 This occurs for a rather well defined value of the energy barrier
  $\Delta \Omega  /k_B T \propto {\rm ln}(Lt/b\tau )$.
  Neglecting the variation of the microscopic
  prefactors $b$ and $\tau$ with temperature and pore size,
  the theory and the experiment will be in good agreement if \\
i) the model reproduces the experimental value of the extrusion
pressure for various temperatures and pore sizes using a single
value of $\Delta \Omega /k_BT$,  and\\
ii) the numerical value of
$e^{-\Delta \Omega/k_BT}$ lies in the range of the expected order
of magnitude of the prefactor  $ b\tau/Lt$.

\begin{figure}
\includegraphics[width=12cm]{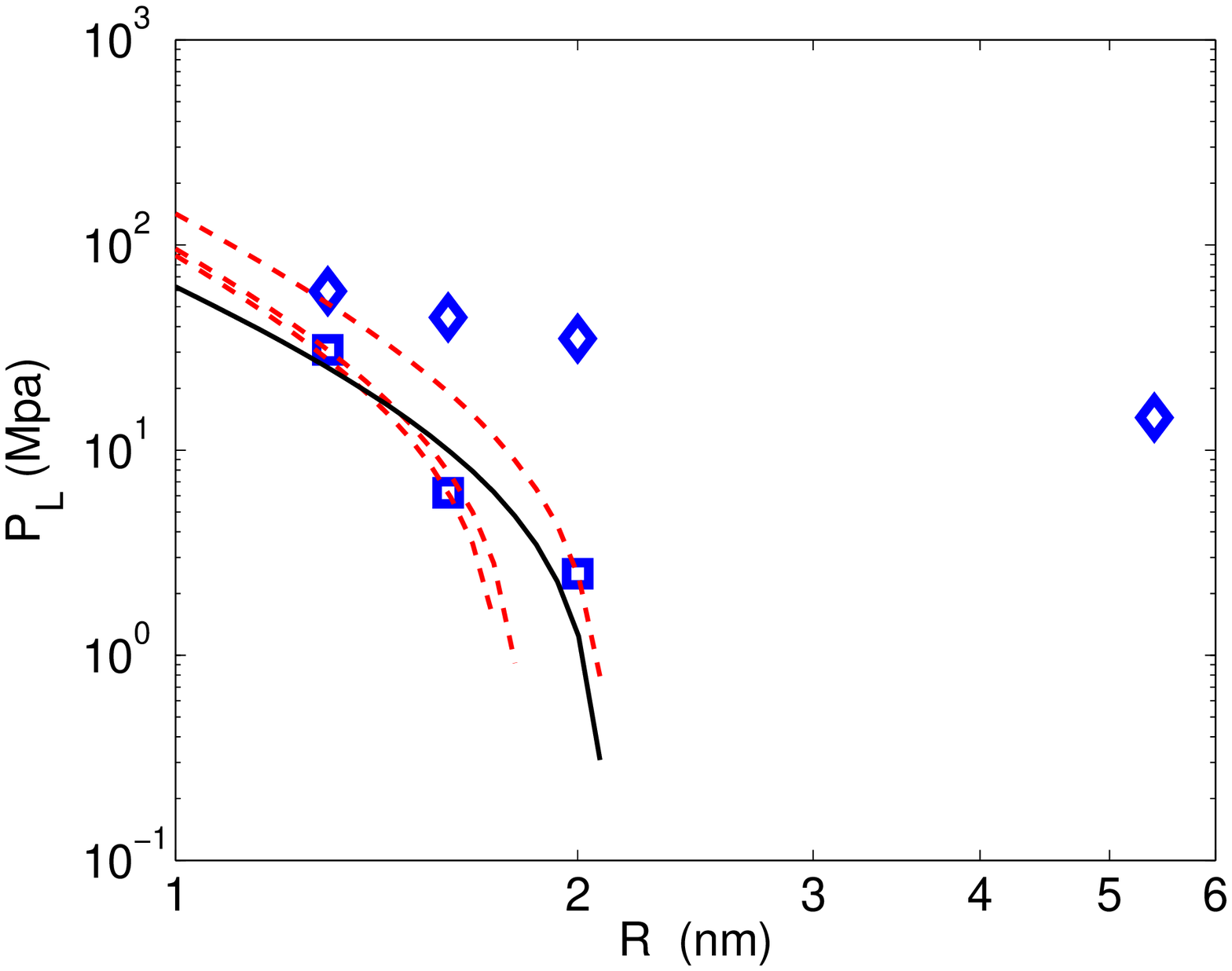}
  \caption{Extrusion (squares) and intrusion (diamonds)  pressure versus pore radius.
   Dotted lines: theoretical curves for the extrusion obtained using
  equation \ref{eqn:approx}, and assuming values of
  $\Delta \Omega= 135k_BT$, $\Delta \Omega= 142k_BT$
  and $\Delta\Omega=190 k_BT$ for the extrusion process.
  Full line: theoretical curve obtained using
  equation \ref{linetension}, with $\Delta \Omega=35k_BT$ and
  $\lambda=-2.4\times 10^{-11}\ \mathrm{J.m}^{-1}$.
  }\label{results}
\end{figure}

In figure \ref{results}  the dotted lines show the plots of the
theoretical extrusion pressure as a function of the pore size
obtained in our model by assuming that extrusion takes place at a
constant value of $\Delta \Omega$. In this calculation we have
used equation \ref{eqn:approx} to obtain the value of $\Delta p$
that results in the desired $\Delta \Omega$, assuming $\theta =
120^o$ and $T=T_1=298K$.
One sees that for a given value of $\Delta \Omega$, the
theoretical extrusion pressure decreases extremely quickly with
increasing pore radius. This variation is much more rapid than for
the  intrusion pressure, and is even stronger  than the one
observed  experimentally for the extrusion pressure. The model
thus provides the correct qualitative tendency but no quantitative
agreement with experiment. The disagreement can be quantified  by
computing the values of the energy barrier $\Delta \Omega$ needed
to provide the measured values of the extrusion pressure. Using
equation \ref{eqn:approx}  we have calculated $\Delta\Omega$ for
each experimental extrusion data point ($P_{ext}^{m}$,
$R_{muff}^{g}$). The resulting values of
   $\Delta\Omega$ are $142\, k_{B}T_{1}$, $135\, k_{B}T_{1}$ and $190\, k_{B}T_{1}$ for
samples MTS-1g, MTS-2g and MTS-3g respectively.  It is seen that
that the nucleation model is much better than the Laplace-Washburn
law for describing the extrusion process : the relative variation
of the energy barrier obtained for the different data points is
only 40\% whereas the extrusion pressure varies by one order of
magnitude. However the quantitative agreement is not perfect.
There is a tendency for the calculated energy barrier to increase
with the pore size, which means that the model overestimates the
energy barrier for large pores.

It is of interest to point out that this overestimation is present
only for the pore size variation of the extrusion pressure. Its
variation with temperature seems very well described by the
nucleation model. We  have calculated the energy barrier of
nucleation  for the  experimental data point corresponding to the
sample MTS-1g at $T_{2}=323 K$ ($P_{ext}^{m} = 37.8\, MPa,\,\,
R_{muff}^{g} = 1.3\, nm $), taking into account the tabulated
value of  water surface tension at $T_{2}=323 K$ and assuming that
the equilibrium contact angle $\theta$ remains unchanged
($\theta=120^o$). We find $\Delta\Omega_{c}=142 k_{B}T_{2}$. We
have checked that changing the value of $\theta$ in a range
consistent with the intrusion pressure does not modify
significantly this energy barrier. This excellent agreement with
the value obtained at $T_{1} = 298 K$ confirms that the extrusion
mechanism in the sample under study is indeed governed by a
thermally activated nucleation process.

However, besides failing to describe precisely the pore radius
dependency of the extrusion pressure, another drawback of the
simple nucleation  model expressed by equations \ref{eqn:barred}
and \ref{nucleation}  is the  high value found for the activation
energies. The numerical value of the exponential factor
$e^{-\Delta \Omega/k_BT}$ found is about $10^{-60}$. Taking for
microscopic parameters a molecular size $b=1 \AA$ and a typical
time between molecular collisions $\tau=10^{-12}$ s, and for
macroscopic parameters an average length of $100$nm for the
capillaries and a value of $10$ s for the experimental time, one
hardly reaches the order of magnitude $10^{19}$ for the prefactor
entering the probability of nucleation. Therefore nucleation
should not be observable experimentally for the activation
energies found with the calculation.

This is a serious drawback. Of course the range of size under
study is close to the nanometer and the use of plain classical
capillarity at this length scale  may seem a too rough approach.
It is  however puzzling that classical capillarity works so well
to describe the intrusion pressure, and fails to provide
reasonable values for the activation energy controlling the
extrusion pressure. The fact that the experimental intrusion
pressure scales as the inverse of the pore radius shows that the
Laplace low of capillarity is not affected by the nanometric range
of the pore size. The most natural way to improve the  model is
then to introduce another macroscopic parameter which has been
neglected, i.e. the tension of the three-phase contact line
between the free surface of the nucleus and the solid wall. Line
tension is a macroscopic thermodynamic parameter similar to
surface tension, whose microscopic source is the modification of
molecular interaction energies close to a three-phase line. These
interactions  induce a deviation of the liquid-vapor interface
from the macroscopic shape it would have if only surface tension
were taken into account, and results in an excess energy which
scales extensively with the system size i.e. with the length of
the contact line. Line tension effects are usually neglected in
classical capillarity when dealing with large systems. A crude
argument for this is that the ratio of line tension to surface
tension has the dimension of a length, and can only be of the
order of a molecular size. Hence line tension will influence only
objects of size comparable to molecular sizes. Based on this
argument, an order of magnitude  for  the line tension can also be
constructed by dividing a  cohesion energy by a molecular size,
yielding $\lambda \simeq k_BT/b \simeq 10^{-11}J/m$.

For nucleation problems which involve  an object of  size
comparable to  Kelvin's radius, line tension effects may however
become important. Oxtoby et al \cite{talanquer} have shown indeed
that incorporating line tension effects improve significantly
macroscopic models  for the nucleation in slit pores.

It is important to mention that line tension effects should not
change the value of the intrusion pressure in the case of a
cylindrical capillary. The reason is that in that particular
geometry, the existence of a line energy does not change the
contact angle of a spherical-cup meniscus with the solid wall.
Therefore, in the case of a model porous medium like MCM41, line
tension is expected to affect only the free energy of the critical
nucleus and the value of the extrusion pressure, but not the
intrusion pressure.

Since there are very few data available for the values of line
tension, we choose to treat it as an adjustable parameter
$\lambda$. For the simplicity of calculations, we make use of two
extra hypothesis for incorporating line tension in our model. The
first one is to assume that the shape of the critical nucleus does
not depend on the value of the line tension. This amounts to
neglect the variation of the contact angle with the line tension.
This variation would be a second order effect here and can be
neglected in a first estimation. With this first hypothesis,
incorporating the line tension contribution to the energy barrier
is very simple. One adds to the energy barrier calculated in
section \ref{theory} the free energy $\lambda l$ of the
three-phase line of the critical nucleus. The length $l$ of this
three-phase line can be calculated from the shape of the critical
nucleus found in section \ref{theory}. We find that its reduced
value $l/R$ varies slightly with the pore radius $R$ (see figure
\ref{errorplotcomp}, which gathers the maximum range of variation
of the reduced length $l/R$ of the critical nucleus contact line
for several values of the contact angle). As a second hypothesis,
we neglect this variation and consider that the length $l$ of the
contact line entering in the energy barrier is $l=K_{3}(s) R$,
where the constant $K_{3}(s)$ depends only on the contact angle
$\theta$ and on the shape of the nucleus $(s)$. The values of
$K_{3}(s)$ are listed in table \ref{tablex}, as well as the
maximum relative error induced by this assumption. Using
approximation \ref{eqn:approx}, the energy barrier for nucleation
in presence of line tension effects writes simply :
\begin{equation}
\Delta \Omega_c = P_L K_1(s) R^3+ \gamma_{lv} K_{2}(s) R^2 +
\lambda K_3(s) R \label{linetension}
\end{equation}
The procedure we use to compare this expression to the
experimental data is to treat $\Delta\Omega_{ext}$ (the value of
the barrier for which extrusion is observed)  and $\lambda$ as two
adjustable parameters and fit equation \ref{linetension} against
the three experimental data  points ($P_{ext},R$) obtained for the
three samples in which extrusion is observed. We then plot in
figure \ref{results} the curve $P_{ext}(R)$ derived from equation
\ref{linetension} with the value of $\Delta\Omega$ and  $\lambda$
obtained from the fit. This curve represents the theoretical
prediction for the extrusion pressure, which takes place for each
sample at the same value of the activation energy. One can see
that the nucleation model with line tension accounts reasonably
well for experimental data, considering the approximations made.
The optimal value for the line tension is  $\lambda =-(2.4\pm
0.3)10^{-11}J/m$, which is a realistic order of magnitude
\cite{herminghaus}. The sign of this tension is negative, which
explains that nucleation occurs relatively more easily in small
pores than expected. We have no explanation  based on microscopic
arguments on why $\lambda$ should be negative in this particular
system, but we note that negative $\lambda$ have already been
reported previously \cite{talanquer,herminghaus}. Finally, the
value found for the energy barrier with the line tension effect is
$\Delta \Omega= (35\pm 5)k_BT$. This value is quite reasonable.
Using as above the estimates  $\tau=10^{-12}$s and $b=1\AA$ for
the microscopic parameters in equation \ref{nucleation},  a
probability of nucleation equal to 1 in a channel of length
$100$nm is obtained after a macroscopic time of $t=1 s$.

\section{Conclusion}

In this paper, we have presented a combination of experimental and
theoretical approaches to describe the intrusion/extrusion cycle
of a nonwetting liquid (water) in an "ideal" hydrophobic porous
material, made of parallel cylindrical pores. The situation is
somewhat similar to the more standard case of capillary
condensation of a wetting liquid in a porous material, with
intrusion taking place at a pressure higher than the bulk
coexistence pressure.

The intrusion data can be well accounted for by using standard
capillarity theory, with an intrusion pressure that scales as the
inverse of the pore radius.  The advancing contact angle obtained
is close to  $120^o$, which is perfectly reasonable for this type
of materials, and suggests that the intrusion branch is close to
equilibrium.

The description of the extrusion process is somewhat more
difficult. Our data suggests that a description similar to that of
the intrusion branch, but using  a receding contact angle, is not
appropriate. Rather, it appears that the limiting step in the
extrusion process is the nucleation of the wetting phase (here the
vapour) that we describe using standard nucleation theory. The
free energy barrier for nucleation is calculated using a standard
capillary description, allowing for various shapes of the critical
nucleus. The experimental data is analyzed by assuming that
extrusion occurs at a fixed value (relative to $k_BT$) of the
energy barrier. This allows a quantitative comparison between
predicted and measured extrusion pressures, as a function of pore
radius. When line tension effects are neglected, the theory cannot
account for the experimental data, and predicts nucleation
barriers that are far too large. However, inclusion of a negative
line tension as an adjustable parameter results in an excellent
agreement between nucleation theory and experiment. The optimal
value of the line tension is in reasonable agreement -as far as
order of magnitude is concerned- consistent with values obtained
using completely different approaches \cite{herminghaus}. Our work
confirms the importance of line tension effects on some aspects of
capillary phenomena at the nanometer scale, and provides an
independent estimate of the line tension.

Two predictions of the present model remain to be tested
experimentally. According to figure \ref{results}, the extrusion
pressure should vanish for pores of radius larger than 2-3nm. This
is consistent with our result that no extrusion is observed for a
sample with $R=4$nm, but samples with intermediate radius values
would be needed to confirm the theory. On the small pores side,
our calculation predicts that extrusion pressure should equal
intrusion pressure for $R=9\AA$. Hysteresis should vanish for such
small pores, for which however the validity of the macroscopic
analysis may become questionable.

 \vskip 1cm

{\bf Acknowledgements}

We are indebted to T. Martin, C. Biolley, A. Galarneau, D. Brunel,
F. Di Renzo, F. Fajula (Laboratoire des Mat\'eriaux Catalytiques
et Catalyse en Chimie Organique, Montpellier, France) for
providing us with the materials and for fruitful discussions as
well as to R. Denoyel (Centre de Thermodynamique et
Microcalorim\'etrie, Marseille, France). We also thank G. Thollet
(Groupe d'Etudes en M\'etallurgie
   Physique et Physique des Mat\'eriaux, Lyon, France) for
    TEM images. B. Lefevre and A. Saugey are supported by
    the French Ministry of Defense (DGA).

\bibliographystyle{unsrt}
\bibliography{jcp}


\end{document}